\documentclass[reprint,aps,nofootinbib,preprintnumbers,amsmath,amssymb,11pt]{revtex4}
\usepackage{graphicx}
\usepackage{dcolumn}
\usepackage{bm}
\usepackage{natbib}
\usepackage{color}
\newcommand{\nc}{\newcommand}
\newcommand\fft[2]{\frac{#1}{#2}}
\newcommand\ft[2]{{\textstyle\frac{#1}{#2}}}
\newcommand\nn{{\nonumber}}
\newcommand{\beq}{\begin{equation}}
\newcommand{\eq}{\end{equation}}

\newcommand{\etas}{\frac{\eta}{s}}
\newcommand{\half}{\frac{1}{2}}

\nc{\bea}{\begin{eqnarray}} \nc{\ea}{\end{eqnarray}} \nc{\be}{\begin{equation}} \nc{\ee}{\end{equation}} \nc{\barr}{\begin{array}}
\nc{\earr}{\end{array}}

\def\be{\begin{equation}}
\def\ee{\end{equation}}
\def\bea{\begin{eqnarray}}

\def\eea{\end{eqnarray}}

\def\l{\lambda}
\def\lab{\label}
\def\f{\Phi}

\def\le{\left}
\def\ri{\right}

\def\half{\frac12}

\def\cO{{\cal O}}

\def\6{\partial}

\def\a{\alpha}
\def\b{\beta}
\def\lab{\label}

\def\g{\gamma}
\def\tr{{\rm Tr}}
\newcommand{\bom}[1]{\fboxsep2mm\fbox{
           $ \displaystyle{ #1} $}}

\begin{document}

\title{On the Temperature Dependence of the Shear Viscosity and Holography}
\preprint{DAMTP-2012-45 \; MIFPA-12-22}

\author{ Sera Cremonini$ ^{\,\clubsuit,\spadesuit}{}^*$, Umut G\"ursoy$^{\dagger}$ and Phillip Szepietowski$ ^{\diamondsuit}{}^\ddagger$
\\[0.4cm]
\it $ ^\clubsuit$ Centre for Theoretical Cosmology, DAMTP, CMS,\\
\it University of Cambridge, Wilberforce Road, Cambridge, CB3 0WA, UK \\ [.5em]
\it $ ^\spadesuit$ George and Cynthia Mitchell Institute for Fundamental Physics and Astronomy\\
\it Texas A\&M University, College Station, TX 77843--4242, USA \\ [.5em]
\it $\dagger$ \textit{Theory Group, Physics Department, CERN, CH-1211 Geneva 23, Switzerland}\\
[.5em]
\it $ ^\diamondsuit$\textit{Department of Physics, University of Virginia,\\ Box 400714, Charlottesville, VA 22904, USA}}
{\let\thefootnote\relax\footnotetext{$^{*}$sera@physics.tamu.edu \\
$^{\dagger}$umut.gursoy@cern.ch \\ $^{\ddagger}$pgs8b@virginia.edu}}

\date{\today}

\begin{abstract}
We examine the structure of the shear viscosity to entropy density ratio $\etas$ in holographic theories of gravity coupled to a scalar field,
in the presence of higher derivative corrections.
Thanks to a non-trivial scalar field profile, $\etas$ in this setup generically runs as a function of
temperature.
In particular, its temperature behavior is dictated by the shape of the scalar potential
and of the scalar couplings to the higher derivative terms.
We consider a number of dilatonic setups, but focus mostly on phenomenological models that are QCD-like.
We determine the geometric conditions needed to identify local and global minima for $\etas$ as a function of temperature,
which translate to restrictions on the signs and ranges of the higher derivative couplings.
Finally, such restrictions lead to an holographic argument for the existence of a global minimum for $\etas$ in these models,
at or above the deconfinement transition.
\end{abstract}

\maketitle
\newpage
\tableofcontents
\newpage

\section{Introduction}

Over the past decade holography has emerged as a valuable tool for gaining insight
into the physics of strongly coupled gauge theories.
Top-down studies based on string/M-theory setups have been met by a number of bottom-up constructions,
with applications ranging from the realm of quantum chromodynamics (QCD) to that of condensed matter systems.
Holographic techniques have been particularly useful for probing the hydrodynamic regime of strongly interacting thermal field theories,
a regime which is notoriously difficult to study directly and -- unlike thermodynamics -- poses a challenge to lattice simulations.

Within this program, many efforts have been directed at better understanding the dynamics of the strongly coupled
QCD quark-gluon plasma (QGP), and in particular at computing its transport coefficients.
One of the most exciting results which has emerged from the heavy ion program at RHIC -- and now at LHC --
is the observation that the hot and dense nuclear matter produced in the experiments displays collective motion.
In fact, the QGP fireball created in off-central collisions is not azimuthally symmetric, but rather shaped like an ellipse.
As a result, the pressure gradients between its center and its edges vary with angle, giving rise to an anisotropic particle distribution.
The matter formed in the collisions then responds as a \emph{strongly coupled fluid} to the differences in these pressure gradients,
displaying a collective flow which is well described by nearly ideal hydrodynamics
with a very small ratio of shear viscosity to entropy density $\etas$.

Experimentally, the flow pattern can be quantified by Fourier decomposing the particles' angular distribution.
In particular, it is the second Fourier component $v_2$, the so-called \emph{elliptic flow}, which is the largest in non-central
collisions and is the observable most directly tied to the shear viscosity.
Thus, bounds on $\etas$ can be extracted from elliptic flow measurements,
with the most advanced analysis at the moment giving $ 4 \pi \, \etas \leq 2.5$ \cite{Song:2010mg}.
Higher order harmonics -- initially neglected because they were assumed to be too small for symmetry reasons --
also play an important role in determining the shear viscosity (see Section IV for a more detailed discussion).
We refer the reader to \cite{Teaney:2003kp,Luzum:2008cw} for some early references on the
RHIC results and the range of $\etas$, and to \cite{Luzum:2010ag,Nagle:2011uz,Shen:2011eg,Muller:2012zq}
for more recent ones including discussions of the first LHC results.

A remarkable result that has emerged from holographic studies of strongly coupled gauge theories
has been the \emph{universality} of the shear viscosity to entropy ratio \cite{Policastro:2001yc,Buchel:2003tz},
which was shown to take on the particularly simple form $\etas=\frac{1}{4\pi}$ in any gauge theory plasma
with an Einstein gravity dual description\footnote{An exception is the case of anisotropic fluids, as first observed in \cite{Erdmenger:2010xm}.}.
Its order of magnitude agreement with RHIC (and now LHC) data was one of the driving motivations behind the efforts to
apply holography to the transport properties of the QGP (see \cite{CasalderreySolana:2011us,Adams:2012th} for recent reviews).
It is by now well understood that deviations from the universal result $\etas=\frac{1}{4\pi}$ (both below and above)
are generic once curvature corrections to the leading Einstein action are included\footnote{For reviews of
the shear viscosity bound
we refer the reader to \cite{Sinha:2009ev,Cremonini:2011iq,Banerjee:2011tg}.}.
Moreover, when there is another scale $\tilde{\Lambda}$ in the system in addition to temperature $T$,  the viscosity to entropy ratio
typically runs as a function of $T/\tilde{\Lambda}$ in such higher derivative theories.
We should emphasize that this type of temperature flow for the shear viscosity arises in a number of holographic constructions,
from theories of higher derivatives in the presence of a chemical potential \cite{Cai:2008ph,Myers:2009ij,Cremonini:2009sy}
or non-trivial scalar field profiles\footnote{Dilatonic couplings to higher derivative terms
in the context of $\etas$ have also been studied in \cite{Cai:2009zv}.}
\cite{Cremonini:2011ej,Buchel:2010wf} and also to systems with spatial
anisotropy \cite{Ghodsi:2009hg,Erdmenger:2010xm,Basu:2011tt,Erdmenger:2011tj,Rebhan:2011vd,Oh:2012zu,Mamo:2012sy}.

\begin{figure}[h!]
 \begin{center}
\includegraphics[scale=0.5]{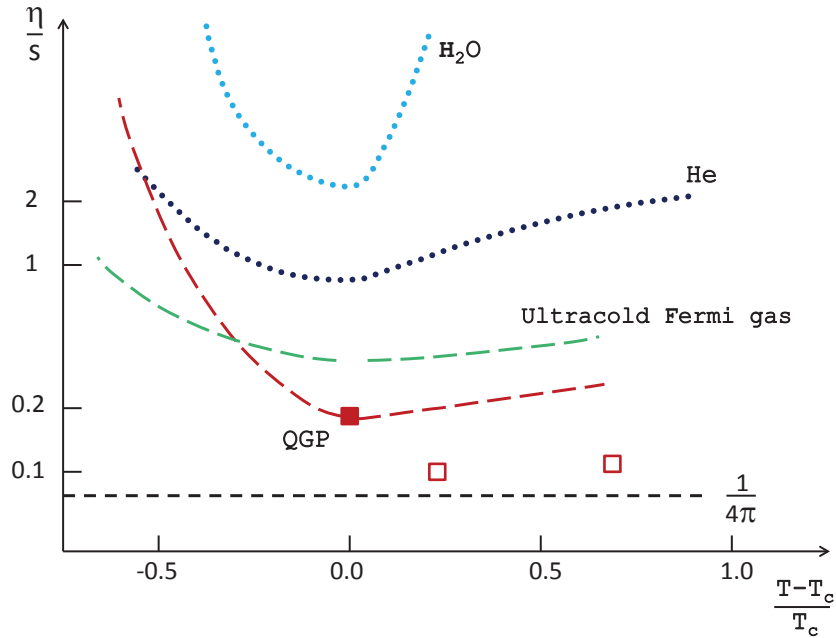}
 \end{center}
 \caption[]{Schematic plots of the shear viscosity to entropy density ratio for a number of fluids in nature.
 $T_c$ denotes the critical temperature at the endpoint of the liquid-gas transition for water and helium,
 the superfluid transition temperature for ultracold Fermi gases and the deconfinement temperature for QCD.
For the water and helium data (dotted lines) we refer the reader to \cite{WaterHelium}.
The dashed curves are the expected theoretical curves for QCD (red) and ultracold Fermi gases (green), from
\cite{Arnold:2000dr,Prakash:1993bt} and \cite{Massignan:2005zz,Mannarelli:2012su} respectively.
%
%
The solid red square denotes the upper bound $\etas \sim 2.5 \left(\frac{1}{4\pi}\right)$ for the QCD quark gluon plasma found in \cite{Song:2011hk},
while the open red squares denote the upper limits found in the lattice analysis of \cite{Meyer:2007ic}.
The dashed horizontal line is the universal ratio $\etas = \frac{1}{4\pi}$.
Similar plots can be found in \cite{Adams:2012th,Lacey:2006bc}.}
\label{EtasComparison}
\end{figure}

The viscosity to entropy ratio is in fact known to be temperature dependent for a variety of liquids and gases in nature (as well as
for ultracold fermionic systems close to the unitarity bound),
exhibiting a minimum in the vicinity of a phase transition (see Figure \ref{EtasComparison}).
A similar behavior is expected \cite{Csernai:2006zz} for $\etas$ near the temperature $T=T_c$ of the
QCD phase transition which separates hadronic matter from the QGP phase.
In the hadronic phase below $T_c$, the hadronic cross section decreases as the temperature is lowered,
leading to an increase in $\etas$ \cite{Gavin:1985ph,Prakash:1993bt}
(for an analysis of transport in the hadronic phase see e.g. \cite{NoronhaHostler:2008ju}).
On the other hand, in the deconfined phase at temperatures well above $T_c$, asymptotic freedom dictates that $\etas$ should increase
with temperature (the coupling between quarks and gluons decreases logarithmically \cite{Arnold:2000dr,Arnold:2003zc}).
From the behavior in these two opposite regimes, we conclude that
we should expect a minimum for $\etas$ somewhere in the intermediate range.

A precise determination of the temperature behavior of transport coefficients such as $\etas$ is an important ingredient
for understanding the dynamics of the strongly coupled medium produced at LHC and RHIC, and may also help in finding the location
of the critical point.
However, at the moment most hydrodynamical simulations of the QGP assume that $\etas$ is a constant, and therefore insensitive to temperature.
The question of the possible relevance of a temperature-dependent $\etas$ on the collective flow of hadrons in heavy ion collisions
has been investigated in a number of studies \cite{Niemi:2011ix,Nagle:2011uz,Shen:2011kn,Bluhm:2010qf},
which thus far have focused mostly on qualitative effects.
The results of \cite{Niemi:2011ix} seem to indicate that at LHC energies elliptic flow values are sensitive to the temperature behavior of $\etas$
in the QGP phase, but insensitive to it in the hadronic phase (with the results reversed at RHIC energies).

Motivated by the potential sensitivity of elliptic flow measurements to \emph{thermal variations} of $\etas$ at the energies probed by LHC,
here we would like to initiate a systematic study of the flow of the shear viscosity as a function of temperature in the context of holography.
In this paper we will restrict our attention to theories with a vanishing chemical
potential, described simply by gravity coupled to a scalar field in the presence of higher derivative corrections.
Although we will not consider string theory embeddings of these backgrounds, the higher derivative terms
we examine are generically expected to show up as corrections to the two-derivative Einstein-scalar action
viewed as an effective field theory, and in particular to most  effective actions derived from string theory.
Moreover, our result for $\etas$ in (\ref{etasG}) is completely generic and applicable to
any top-down model which contains a truncation to an Einstein-scalar theory.

Thanks to the presence of a non-trivial scalar field profile, the higher derivative terms in our theory will generically generate
a temperature flow for the shear viscosity, as already expected from \cite{Cremonini:2011ej,Buchel:2010wf}.
In particular, the temperature dependence of $\etas$ will be dictated by the shape of the scalar potential and of
the couplings of the scalar to the higher derivative interactions.
Given an explicit expression for $\etas$ in terms of the latter, it is then straightforward to discuss the
existence of \emph{local} minima for $\etas$ as a function of temperature.
As we will see, for potentials which confine quarks at zero temperature it is also possible to determine
criteria for the existence of \emph{global} minima.
In particular, the requirements that $\etas$ approaches its high-temperature value from below
(dictated from asymptotic freedom) and that the zero-temperature theory is confining,
are enough to fix the signs (and ranges) of the couplings of the higher derivative terms, and
guarantees the existence of a global minimum for $\etas$, at or above $T_c$.
Thus, we have been able to give a holographic argument for the existence of a minimum for $\etas$
in this class of models, along with a geometric interpretation for it,
complementing existing field-theoretic\footnote{The authors of \cite{Kovtun:2011np} tie a lower bound
on the shear viscosity to the validity of second-order hydrodynamics.} arguments \cite{Kovtun:2011np}.
Finally, although our analysis will be more general, we will focus mostly on `phenomenological' models engineered to
reproduce some of the qualitative and quantitative features of QCD.

The rest of the paper is organized as follows.
Section II describes our setup and outlines how to obtain $\etas$ for higher derivative corrections to
dilatonic black brane solutions.
Our main results are presented in Section III, where we discuss the temperature dependence of $\etas$ in
theories with a non-trivial scalar field profile, first by considering a toy model consisting of an exponential scalar potential,
then focusing on QCD-like phenomenological models.
We also comment on qualitative features of $\etas$ in theories that are confining, and identify generic criteria
for the existence of a minimum for $\etas$ as a function of temperature in such setups.
We summarize our results in Section IV, where we also comment on the challenges involved with determining more precisely the structure
of $\etas$, and in particular how it flows with temperature.

\section{Setup}

In order to generate a non-trivial temperature dependence for $\etas$ at zero chemical potential
we will consider backgrounds with a
scalar field coupled to a higher derivative theory of gravity.
The action we consider is of the form
\bea
\label{fullaction}
S &=& \fft1{16\pi G_5} \int d^{5}x\sqrt{-g} \, \Bigl[R - 2 (\nabla\Phi)^2 + V(\Phi)
+ \ell^2 \beta \, G(\Phi) \, R_{\mu\nu\rho\sigma}R^{\mu\nu\rho\sigma} \Bigr] \, , 
\ea
under the assumption that the coupling $\beta$ of the higher derivative terms is perturbatively small and $G(\Phi)$ is an arbitrary regular function of $\Phi$. For most of the paper we will focus on the exponential case, $G(\Phi) = e^{\gamma\Phi}$, however most of the analysis is straightforwardly generalized to arbitrary $G(\Phi).$
The dilaton potential $V(\Phi)$ is assumed to have either a minimum at some value $\Phi=\Phi_0$ or
a run-away behavior $V(\Phi)\to const$ as $\Phi\to -\infty$ (as in \cite{GK, GKN}).
Although in the above we have only considered the higher derivative correction $R_{\mu\nu\rho\sigma}R^{\mu\nu\rho\sigma}$,
this is in fact the only term which contributes to $\etas$ at this order in the derivative expansion.
The computation of $\etas$ in theories with higher derivatives has been studied in great detail, and
here we review briefly only the relevant aspects\footnote{For reviews of higher derivative corrections to $\etas$ and further
details of the computation we refer interested readers to \cite{Sinha:2009ev,Cremonini:2011iq,Banerjee:2011tg} and references therein.}.

\subsection{Extracting the Shear Viscosity to Entropy Ratio}

In the hydrodynamic approximation to near-equilibrium dynamics, the transport coefficients of a finite temperature
plasma can be extracted in a number of ways.
The most straightforward method for computing the shear viscosity is based on the Kubo relation
\beq
\label{etaGreen}
\eta = - \lim_{\omega\rightarrow 0} \frac{1}{\omega} \, \text{Im} \,  G^R_{xy,xy}(\omega,\vec{k}=0) \, ,
\eq
which reduces $\eta$ to the low frequency and zero momentum limits of the stress tensor's retarded Green's function.
Using the holographic dictionary, the relevant two-point correlation function of the shear stress tensor  $T_{xy}$
can be read off from the effective action of a shear metric fluctuation
$ h_{\,x}^{\;\;\;y} (t,u) \sim \int d^4k \, \phi_k (u) \, e^{-i\omega t + i k z}$,
where we use $u$ to denote the radial direction in the bulk.
Expanding (\ref{fullaction}) to quadratic order in the modes $\phi_k$, one finds the by now standard\footnote{See \cite{Buchel:2004di} for
the original derivation, in the context of $\alpha^{\prime\, 3}$ corrections.} form
of the effective action for the shear fluctuation,
\begin{eqnarray}
\label{effaction}
S_{eff} & \sim \int \frac{d^4k}{(2\pi)^4} \; du & \Bigl[ A(u) \, \phi^{\prime\prime}_k \phi_{-k} +
B(u) \,\phi^{\prime}_k \phi^{\prime}_{-k} +
C(u) \, \phi^{\prime}_k \phi_{-k} + \nn \\ && + \; D(u) \, \phi_k \phi_{-k} + E(u) \, \phi^{\prime\prime}_k \phi^{\prime\prime}_{-k} + F(u) \,\phi^{\prime\prime}_k
\phi^{\prime}_{-k} \Bigr] + S_{GH} \, ,
\end{eqnarray}
where the coefficients $A(u), B(u),\ldots, F(u)$
encode information about the background solution, and
$S_{GH}$ is the generalized Gibbons-Hawking boundary term.

After a number of subtle manipulations, the shear viscosity can be extracted from (\ref{effaction}) and reduces
\cite{Myers:2009ij} to the compact expression
\beq
\eta = \left. \frac{1}{8\pi G_5}
\left[
\sqrt{-\frac{g_{uu}}{g_{tt}}} \left(A-B+\frac{F^{\, \prime}}{2}\right) +
\left(E \left( \sqrt{-\frac{g_{uu}}{g_{tt}}} \, \right)^{\, \prime} \, \right)^{\, \prime} \;
\right] \, \right\vert_{u=u_h} \; ,
\eq
evaluated at the horizon radius $u_h$, showing that it is
given entirely in terms of horizon data\footnote{The Kubo formula (\ref{etaGreen}) can be shown \cite{Iqbal:2008by,Myers:2009ij}
to be equivalent to
\beq
\label{eq:etapi}
\eta = \lim_{u,\omega\rightarrow0} \frac{\Pi_{\omega,k=0}(u)}{i\omega \, \phi_{\omega,k=0}(u)}\, ,
\eq
where $\Pi_k$ is the radial-momentum conjugate to $\phi_k$.
In the low frequency limit (as long as the boundary theory is spatially isotropic) the quantity inside the limit
does not depend on the radial coordinate. As a result, it can be evaluated at an arbitrary value of $u$, and
in particular at the horizon \cite{Iqbal:2008by,Myers:2009ij}.}.
Finally, the entropy density $s$ is easily found by dividing by the (infinite) black brane volume Wald's entropy formula,
\begin{equation}
\label{eq:Wald}
S = - 2 \pi \int_{\Sigma} d^{3} x \sqrt{- h} \frac{\delta {\mathcal L}}{\delta R_{\mu \nu \rho \sigma}} \,
\epsilon_{\mu \nu} \epsilon_{\rho \sigma} \, ,
\end{equation}
where $h$ is the induced metric on the horizon cross section $\Sigma$, and $\epsilon_{\mu \nu}$ the binormal to $\Sigma$.

From our discussion above it is evident that $\etas$ can be expressed entirely in terms of near-horizon data.
Moreover, when working perturbatively in the coupling $\beta$ of the higher derivative terms,
$\etas$ can be determined purely from the background solution of the two-derivative theory\footnote{Because of the universality
of $\etas$, ${\cal O}(\beta)$ corrections to the background geometry
lead to order ${\cal O}(\beta^2)$ corrections to $\etas$.}.
These two facts allow us to write $\etas$ in terms of the parameters of a generic near-horizon non-extremal black brane expansion.
Parametrizing the black-brane solution to the leading order two-derivative action by
\beq
\label{metric1}
ds^2 = -a^2(u) \, dt^2 + c^2(u) \, du^2 + b^2(u) \, d\vec{x}^2 \, , \quad \f = \varphi(u) \, ,
\eq
with the choice $b^2(u) = 1/u$, we can write down its near-horizon expansion by assuming a first order zero in $g_{tt}$
and a corresponding first order pole in $g_{uu}$,
\begin{eqnarray}
\label{aexp}
a(u)^2 &=& a_0(1-u) + a_1(1-u)^2 + a_2(1-u)^3 + ...\,, \nn \\
\label{bexp}
b(u)^2 &=& b_0  (1 + (1-u) + ... )\,, \nn \\
\label{cexp}
c(u)^2 &=& c_0(1-u)^{-1} + c_1 + c_2(1-u) + ...\,, \nn \\
\label{phiexp}
\varphi(u) &=& \varphi_h + \varphi_1(1-u) + \varphi_2(1-u)^2 + ...\; \; .
\end{eqnarray}
The shear viscosity to entropy density ratio is then of the simple form \cite{Cremonini:2011ej}
\begin{equation}
\label{etasNH}
\etas = \frac{1}{4\pi}\left(1 - \frac{\beta \, \ell^2}{c_0} \, (G(\varphi_h) + 2 \varphi_1\, G'(\varphi_h)\,) 
\right)\, ,
\end{equation}
and is only sensitive to the parameters $\{c_0,\varphi_h,\varphi_1\}$ of the near-horizon expansions (\ref{cexp}).
This concludes the derivation of $\etas$ for the theory described by (\ref{fullaction}).
In the remainder of this section we will discuss the class of dilatonic black brane solutions we are interested in.

\subsection{Shear Viscosity of Dilatonic Brane Solutions}

As we mentioned above, the backreaction of the higher derivative terms on the background solution does \emph{not}
affect $\etas$ to linear order in perturbative parameter $\beta$, which is the order that we are interested in here.
Thus, we are going to focus on black brane solutions to models described by the two-derivative action
\bea\label{action0}
S &=& \fft1{16\pi G_5} \int d^{5}x\sqrt{-g} \, \Bigl[R - 2 (\nabla\Phi)^2 + V(\Phi)
 \Bigr] \, ,
\ea
and neglect curvature corrections.
Although analytic black brane solutions are not known for generic choices of the potential,
to extract $\etas$ knowledge of the near-horizon behavior is enough, given our prescription (\ref{etasNH}).
For this calculation we find it more convenient to introduce a new radial coordinate $r$ and parametrize the black brane ansatz as
\begin{equation}
\label{BHuMain}
  ds^2 = f^{-1}(r) \, dr^2 + e^{2A(r)}\le( d\vec{x}^2 - f(r) \, dt^2  \ri), \qquad \Phi =   \Phi(r) \, .
\end{equation}
From the expression (\ref{etasNH}) it is clear that we will need to calculate the near horizon values of the metric functions $c(u)$
and $\varphi(u)$ in (\ref{metric1}), and the relationship between T and $\varphi_h$, which will make the temperature
dependence of (\ref{etasNH}) explicit. We will instead calculate the near horizon values of the functions $\Phi$ and $f$ above and
perform a change of variables at the end to express (\ref{etasNH}) in terms of physical quantities.

For this  study, it turns out to be particularly
convenient to adopt the \emph{phase variables} method developed in \cite{LongThermo}, which we briefly review here
and in Appendix \ref{PhaseVars}. This is a quick and efficient way to obtain the thermodynamic properties of the diatonic branes.
In place of solving the full 5th order set of Einstein's equations, one only needs to solve two first order differential equations
for the so-called ``phase variables'' that are defined from the metric functions as
\begin{equation}\label{XY2}
  X(\f)\equiv \frac{\zeta}{4}\frac{\Phi'}{A'}\, , \qquad  Y(\Phi)\equiv
  \frac{1}{4}\frac{f'}{f\, A'}\, .
\end{equation}
The constant $\zeta$ depends on the normalization of the scalar kinetic term in (\ref{fullaction}) and dimensionality.
In our case it is fixed to be $\zeta=\sqrt{8/3}$. Clearly, these functions are invariant under reparametrizations of the radial coordinate.
Physically, one can interpret the boundary values of these variables as the thermodynamic energy and the enthalpy of the system.
Furthermore, their horizon expansion is completely determined in terms of the
dilation potential $V(\Phi_h)$ and $V'(\Phi_h)$ by the requirement of regularity.
The behavior of the metric functions near the horizon is also determined in terms of these two quantities.
The details of the calculation are explained in Appendix \ref{PhaseVars} and here we present only the final results.

We parametrize the near-horizon expansion of the metric and scalar field of the black brane ansatz (\ref{BHuMain}) at
$r = r_h$ as follows,
\bea
\lab{Ahor}
A(r) &=& A_h + A_1 (r-r_h) + \cdots\\
\lab{fhor}
f(r) &=& f_1 (r-r_h) + \cdots \\
\lab{phihor}
\f(r) &=& \f_h + \f_1 (r-r_h) + \cdots \; .
\eea
We can now make use of the phase variables method to obtain explicit expressions\footnote{In particular, we use
eqs. (\ref{Ap}), (\ref{fp}),  (\ref{XY1}), (\ref{use1}) and (\ref{Teq2}) in Appendix \ref{PhaseVars}.} for
$f_1, A_1$ and $\Phi_1$ in terms of the physical parameters in the system,
\bea
\label{A1f1phi1}
A_1 &=& -\frac{C}{\ell} \frac{S^{\frac13}}{T} V(\f_h) \, , \nn \\
\f_1 &=& \frac{3C}{4\ell} \frac{S^{\frac13}}{T} V'(\f_h)\, , \nn \\
f_1 &=& -M_p (4\pi)^{\frac43} \frac{T}{S^{\frac13}}\, ,
\eea
with $C$ given by (\ref{Cdef}).
As can be seen from (\ref{etasNH}),  the only near-horizon parameters that are needed for extracting $\etas$ are
$c_0$, the leading order coefficient in the expansion of $c(u)$, and  $\varphi_h$ and $\varphi_1$, the first two terms in the scalar field
expansion.
By performing a change of radial coordinate we can relate the two near horizon expansions,
and write $\{ \varphi_h, \varphi_1, c_0 \}$ in terms of $\{A_1,f_1,\f_h,\f_1\}$ as
\beq
\label{parametermap}
\varphi_h = \f_h \, , \quad \quad \varphi_1 = \frac{1}{2A_1} \f_1 \, , \quad \quad c_0 = \frac{1}{2f_1 A_1} \, .
\eq

We now have all the ingredients needed to apply the near-horizon $\etas$ prescription (\ref{etasNH}) to the generic black
brane expansion (\ref{A1f1phi1})--(\ref{parametermap}) we just found.
As expected, we find that the deviation from the universal result $\etas = \frac{1}{4\pi}$ is controlled by the shape of the potential, the horizon
value of the scalar field as well as the coupling $\beta$ of the higher derivative term,
\be
\etas
= \frac{1}{4\pi} \left[1+\frac23\ell^2\beta \left(-G(\f_h)V(\f_h)+ \frac34 G'(\f_h)V^\prime(\f_h)\right)  \right]\, ,
\lab{etasG}
\ee
and for the specific choice of $G(\Phi) = e^{\gamma\Phi},$ which we will focus on for the most of this paper,
\be
\bom{\etas
= \frac{1}{4\pi} \left[1+\frac23\ell^2\beta \left(-V(\f_h)+ \frac34\gamma V^\prime(\f_h)\right)  e^{\gamma \Phi_h} \right]\, . }
\lab{etas}
\ee
We emphasize that this expression is completely general, and applies to any asymptotically AdS solution to (\ref{fullaction}).
Moreover, we can already anticipate that $\etas$ will generically be temperature dependent,
thanks to the presence of a non-trivial scalar field profile, as already seen in \cite{Cremonini:2011ej,Buchel:2010wf}.
To better understand the meaning of (\ref{etas}), we recall that
the flow of $\etas$ as a function of temperature is mirrored, in the bulk, by the change of the
near-horizon geometry of the solution, as the horizon radius varies.
For the class of holographic constructions we are considering here, it is the scalar field profile which is responsible for
introducing an additional scale $\tilde{\Lambda}$ in the theory (in addition to temperature),
and thus breaking the conformal symmetry away from the UV.
As a result, we should think of $\Phi_h$, the horizon value of the dilaton, as tracking the temperature of the system\footnote{More precisely,
it will track the dependence on $T/\tilde{\Lambda}$, where $\tilde{\Lambda}$ is the new scale in the system.}.
Thus,  (\ref{etas})
can be expressed entirely in terms of
temperature by finding the precise relationship between $\Phi_h$ and $T$.
As usual, the latter can be determined from the metric by demanding regularity at the horizon. We review this is Appendix A.

Finally, we would like to point out that in the special case of a non-dynamical scalar field and a constant potential, we recover the
standard result for an AdS black brane in pure gravity with curvature corrections \cite{Brigante:2007nu,Kats:2007mq}, which is well-known to
give rise to a \emph{constant} correction to the universal $\etas = \frac{1}{4\pi}$ result.
More interestingly, there are special choices of non-trivial $V(\f)$ for which the temperature dependence disappears, as we will
see more explicitly below.

\section{Results}

In this section we examine the temperature dependence of $\etas$ in various holographic setups.
We start by looking at a toy model consisting of a simple exponential potential, and continue with more `phenomenological'
constructions designed to mimic QCD and in particular the physics of the strongly coupled quark gluon plasma.
We will conclude by remarking on generic, qualitative features of $\etas$,
including a discussion of the existence of minima as a function of temperature.

\subsection{A Warm-up Example: Chamblin-Reall Black Brane}
\lab{CR}

As a warm-up example, we will work out the {\em Chamblin-Reall} (CR) black-hole solution \cite{CR} by making use of
the phase variables formalism. The CR brane  is a solution to (\ref{action0}) with the single-exponential potential
\be\lab{VCR}
V(\f) = \frac{V_0}{\ell^2} \, e^{\,\a \f},
\ee
where $V_0$ is a positive dimensionless constant and $\ell$ defines a length-scale in the background. We choose $\a>0$.
With this convention\footnote{In the two-derivative theory, the sign of $\a$ can be altered by the transformation $\f\to-\f$.},
in the zero temperature geometry the scalar approaches $\f\to -\infty$ on the boundary and $\f\to+\infty$ in the deep interior.
In the black-brane geometry, which is what we are ultimately interested in, $\f$ runs from $-\infty$ on the boundary to
a constant value $\f_h$ at the horizon. Thus, anywhere outside the horizon $\f<\f_h$.

Although the potential (\ref{VCR}) does not admit an AdS minimum as the scalar field approaches the boundary, the CR brane
-- for a special value of $\alpha$ -- can be obtained from a dimensional
reduction of pure gravity plus a (negative) cosmological constant in six dimensions.
In fact our action (\ref{fullaction}), with the potential choice (\ref{VCR}), can be obtained via
a $U(1)$ reduction from the following six-dimensional mother theory
\begin{equation}
\mathcal{L} =   R + 2\Lambda_{6} + \beta R_{\mu\nu\rho\sigma}R^{\mu\nu\rho\sigma} \, ,
\end{equation}
for the special case\footnote{More general dimensional reductions \cite{Gouteraux:2011qh} may
yield additional values of $\gamma,\alpha$.} of $\gamma = - \alpha = - \sqrt{2/3}$ (see Appendix B for details of the reduction).
This special parameter choice -- for which the CR solution
can be uplifted to a pure AdS black brane in six dimensions  --
will play an interesting role in the behavior of $\etas$, as we will see later in this section.
For now, however, we will keep $\{\alpha,\gamma\}$ completely arbitrary.

The CR solution corresponds to the fixed-point of the X-equation (\ref{Xeq}) with
\be\lab{XCR}
X(\f) = x_0 \equiv -\frac{\a}{2\zeta},
\ee
where recall $\zeta = \sqrt{8/3}.$ Substituting this into the Y-equation (\ref{Yeq}), the solution becomes
\be\lab{YCR}
Y(\f) = \frac{1-x_0^2}{e^{\kappa(\f-\f_h)}-1}, \qquad \kappa \equiv \frac{\zeta}{x_0} (1-x_0^2).
\ee
From (\ref{Aeq1}) and (\ref{feq1}) one can then reconstruct the full background as a function of $\f$,
\be\lab{AfCR}
A(\f) = A_c + \frac{\zeta}{4 x_0} (\f-\f_c)\, , \qquad f(\f) = 1- e^{\kappa(\f_hs-\f)}\, ,
\ee
where $\Phi_c$ corresponds to a cutoff surface, as explained in the appendix.
One can now use (\ref{Teq1}) and (\ref{ent31}) to obtain the entropy and temperature as a function of $\f_h$,
\be\lab{TandS}
T(\f_h) = T_0 \; e^{\frac{\zeta}{4 x_0} (1-4 x_0^2) \f_h},\qquad S(\f_h) = S_0\; e^{\frac{3 \zeta}{4 x_0} \f_h},
\ee
where we defined
\be
\lab{LambdaS0}
T_0 \equiv \frac{V_0}{ 12 \pi \ell } \; e^{A_c - \frac{\zeta}{4 x_0}(1+4x_0^2)\f_c}\, ,
\qquad S_0 \equiv S(0) = \frac{e^{3(A_c-\frac{\zeta}{4x_0}\f_c)}}{4G_N} \,.
\ee
The temperature dependence of the entropy can be read off from (\ref{TandS}),
\be\lab{SF}
S(T) \propto T^{\frac{3}{1-4 x_0^2}} \, .
\ee
Combined with the first law of thermodynamics, this fixes the temperature scaling for the free energy of the system
\be
\lab{FreeEnergyScaling}
F \propto -T^{\frac{4-4 x_0^2}{1-4 x_0^2}} \, .
\ee

In both equations (\ref{SF}) and (\ref{FreeEnergyScaling}) the proportionality constant is positive.
Notice that the free energy $F$ is always negative definite and never crosses zero --
there is no phase transition in the system.
Moreover, the requirement that the specific heat
of the system $C_v = T dS/dT$ is positive definite (for thermal stability) constrains the value of $x_0$, resulting in $x_0^2<1/4$.
Using (\ref{XCR})
this can be translated into a condition on $\a$, the exponent in the dilaton potential (\ref{VCR}),
\be\lab{condCv}
0< \a < \sqrt{\frac{8}{3}}.
\ee
Together with this condition, equations (\ref{SF}) and (\ref{FreeEnergyScaling}) essentially determine
all of the thermodynamic properties of the system.
The fact that -- to work out the thermodynamics associated with this background --
we only needed to solve two first order differential equations, (\ref{Xeq}) and (\ref{Yeq}),
and not the full system of Einstein's equations, demonstrates explicitly the advantage of using the phase variables method.

We are now ready to calculate the shear viscosity of the CR black-brane solution.
Although we are particularly interested in the case of $\gamma = -\alpha = -\sqrt{2/3}$, for which
the CR solution comes from a $U(1)$ reduction of a pure AdS black brane in six dimensions,
here we write down a `formal'\footnote{Although our near-horizon $\etas$ prescription was obtained specifically
for geometries that are asymptotically AdS,
it may be possible to generalize it to backgrounds that contain asymptotically conformally flat radial slices,
by appropriately taking into account holographic renormalization in such backgrounds as in \cite{Wiseman:2008qa}.}
expression for $\etas$ for arbitrary values of $\alpha$ and $\gamma$.
Using (\ref{VCR}) and the corresponding expressions for the temperature and entropy (\ref{TandS}),
one finds
\be\lab{etaSCR}
 \etas = \frac{1}{4\pi}\left[1 - \frac{2}{3}\beta V_0 \left(1-\frac{3\gamma\a}{4}\right) \left(\frac{T}{T_0}\right) ^{\frac{-6\a(\gamma+\a)}{8-3\a^2}} \right] \, .
\ee
Recall from (\ref{condCv}) that $\a^2<8/3$, and therefore the sign of the power of the temperature in (\ref{etaSCR})
depends on whether $\gamma <-\a$ or not.
In either case, $\etas$ is a monotonic function of the temperature, and whether it decreases or increases relative
to $\frac{1}{4\pi}$ depends on the sign of $\beta$ as well as the range of $\gamma$.
Furthermore, if we require the zero temperature limit to
approach the universal $\frac{1}{4\pi}$ result, we must impose $\gamma < - \alpha$.

Interestingly, the temperature dependence of the shear viscosity to
entropy ratio disappears in the two special cases:
\begin{enumerate}
\item $V \propto e^{\frac{4}{3\gamma}\f}$.
In this case not only the $T$-dependence disappears,
but also $\etas$ resumes its universal value $\frac{1}{4\pi}$, despite the presence of higher derivative corrections.
\item $ V \propto e^{-\gamma\f}$.
For the special case of $\alpha=\sqrt{2/3},$ $\etas$ takes exactly the same value as in the six-dimensional AdS
Schwarzschild black hole. This fact can be understood by reducing the $AdS_6$ Schwarzschild solution on
$S^1$.  It inherits the scale symmetry of the parent solution in six-dimensions,
hence leading to absence of the $T$-dependence. See Appendix \ref{app:U1} for details of this calculation. Similar statements can be made for reductions from $(d+1+n)$-dimensions on an $n$-torus as in \cite{Gubser1}, which for $d=4$ would yield  $\alpha = -\gamma = \sqrt{\frac{8n}{3(n+3)}}$.
 \end{enumerate}
It is possible that the cancelation of the correction to $\frac{1}{4\pi}$ in the former case
may also be understood in terms of a dimensional reduction of a parent theory, but this time without a cosmological constant\footnote{We thank
Blaise Gouteraux for a very interesting discussion on this point.}.
At the moment we don't have a more complete understanding of this case (see however \cite{Blaise}).

\subsection{Improved Holographic QCD}\label{ihqcd}

\subsubsection{ihQCD Background}
Next, we turn to the phenomenological models discussed in \cite{GK, GKN, Gubser1} and focus in particular
on the setup of \cite{GKMN1, LongThermo}.
These are phenomenological constructions in the sense that the potential $V(\f)$ is determined purely by field-theoretic requirements,
and is designed to capture some of the features of QCD while remaining reasonably tractable.
In fact, in these models the dilatonic scalar
can be identified\footnote{Up to a multiplicative factor which does not affect physical observables.}
with the running 't Hooft coupling, $\lambda = N_c g_{YM}^2\sim e^\f$, and the scalar potential $V(\f)$ is
directly related to the $\beta$-function of the system, giving a holographic definition of the latter in terms of the background geometry.
Thanks to this identification one can extract the UV and IR asymptotics of the potential
from the small $\lambda$ and large $\lambda$ expansions of $\beta(\lambda)$.

In the UV (small $\l$), the input for the behavior of $V(\f)$ comes from perturbative QCD, \emph{i.e.}
from the requirement of asymptotic freedom with a logarithmic running coupling.
More generally, one assumes that there is a dimension-four operator $\tr F^2$ in the spectrum, dual to the dilaton in the bulk.
The fact that the operator $\tr\, F^2$ is marginal in the UV then translates into the statement that the UV geometry is asymptotically $AdS_5$,
with logarithmic corrections. The details of the UV expansion for these types of backgrounds can be found in \cite{GK}.

On the other hand, in the IR (large $\l$) the potential is fixed by demanding linear quark confinement.
On the bulk side this is implemented by adding a probe string in the geometry -- dual to a Wilson-loop --
and obtaining the quark-anti quark potential from the asymptotics of the string embedding in the IR \cite{Wilson, GKN}.
One finds that linear confinement requires the dilaton potential to have an asymptotic expansion in the IR of the form
\be\lab{IRpot}
V(\Phi) \to e^{Q\Phi}\;  \Phi^P + \cdots \, ,
\ee
where we show only the leading large $\Phi$
behavior\footnote{The difference between this definition and the original one given in \cite{GK}
arises from different normalizations of the dilaton kinetic term.}, with $\sqrt{8/3} \leq Q \leq \sqrt{16/3}$ and $P>0$.
The parameter choice which fits the available zero and finite temperature data best turns out to be $Q=\sqrt{8/3}$ and $P=1/2$ (see \cite{GKMN3}).
This particular choice is also well motivated by the fact that it exhibits desirable
qualitative features such as a linear glueball spectrum,
screening of the magnetic quarks \cite{GKN}, and more recently the scaling behavior of the interaction measure
in temperature \cite{Delta}\footnote{See \cite{Veschgini:2010ws}
and references therein for a criticism of ihQCD models.}.

More specifically, defining $\l = \exp(\sqrt{3/2} \Phi)$,
an example of a potential with the correct UV and IR asymptotics is of the form
\be\lab{pot}
{\ell^2 V(\l) \over 12}= 1+ \l + V_1 \l^{\frac43} \left[\log(1+V_2 \l^{4 \over 3} +V_3 \l^2)\right]^{1/2} \, ,
\ee
where the value at $\l =0$ sets the UV AdS scale $\ell$.
The remaining parameters in the potential (\ref{pot}) can be fixed by matching the scheme-independent $\beta$-function
coefficients of large N QCD, the lowest glue ball mass and the latent heat at $T_c$ \cite{GKMN3}.
Although we will analyze explicitly the model with the potential given in (\ref{pot}), we emphasize that
{\em our qualitative results will only depend on the fact that $V\to const$ in the UV and that it is a confining potential in the IR}.

\subsubsection{Thermodynamics}
\lab{thermo}

\begin{figure}[h!]
 \begin{center}
\includegraphics[scale=0.9]{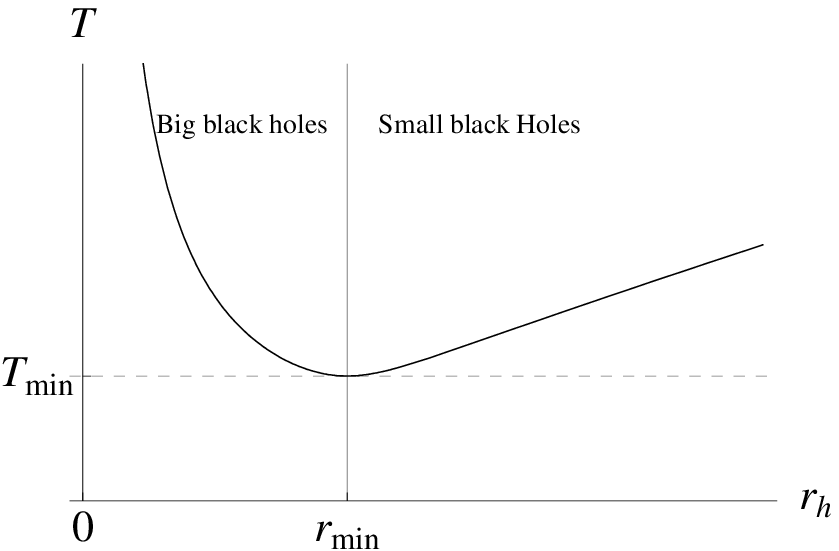}\\
(a)\\
\vspace{1cm}
\includegraphics[scale=0.9]{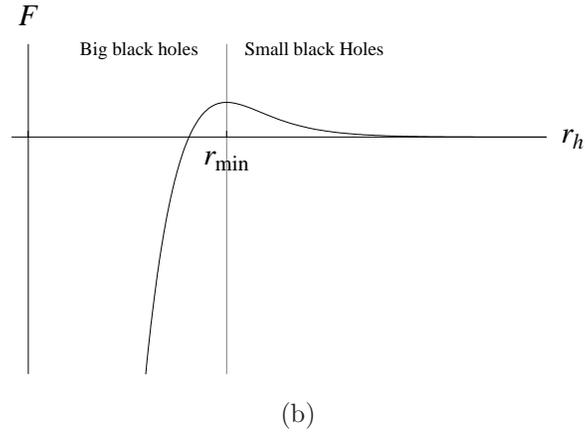}\\
(b)
 \end{center}
 \caption[]{Typical plots of the black-hole temperature (a) and free energy (b)
 as a function of the  horizon position $r_h$, in a confining background.
The temperature features a minimum at $r_{min}$, which
separates the large black-hole from the small black-hole branches.
The locus given by $F(r_c) = 0$, with $r_c < r_{min}$, corresponds to the phase transition point, $T=T_c$.}
\label{fig1}
\end{figure}
Before discussing the temperature dependence of the shear viscosity in this setup, we would like to summarize
the basic thermodynamic properties of the system.
These models exhibit a first order confinement-deconfinement transition at some critical temperature $T=T_c$.
Below $T_c$, the dominant phase is the confined phase that corresponds to a thermal graviton gas background.
On the other hand, above $T_c$ one has the deconfined phase corresponding to a \emph{big} black-hole background.
We note that there also exists a third phase above a certain temperature $T_{min}$, where $T_{min}<T_c$, which
is sometimes referred to as a \emph{small} black-hole. While the horizon of the latter is deep in the interior,
the big black hole has its horizon closer to the boundary.

\begin{figure}[h!]
 \begin{center}
\includegraphics[scale=0.5]{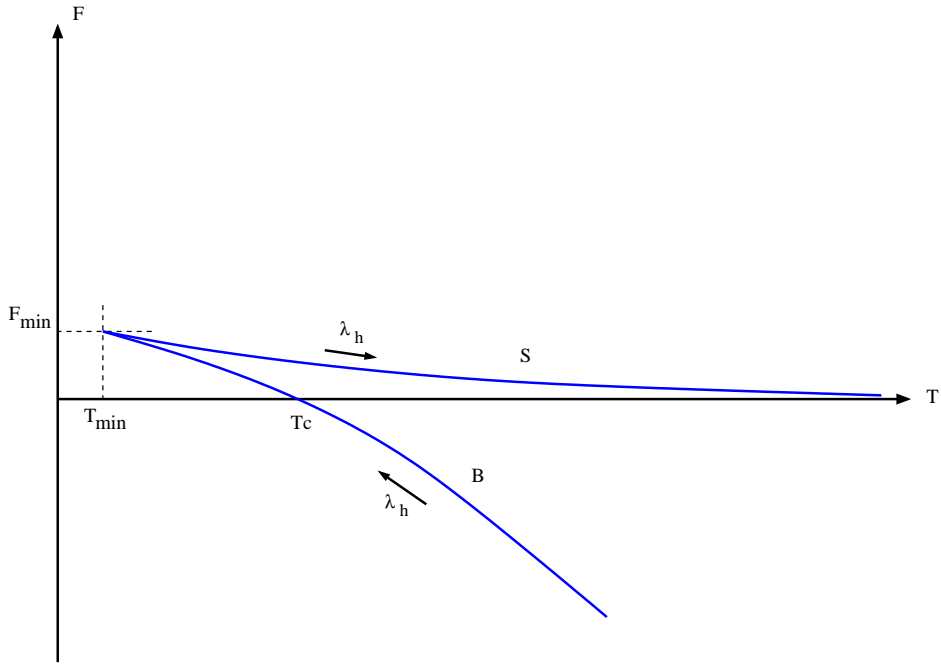}
 \end{center}
 \caption[]{Black hole free energy as a function of temperature.
 The lines ``B'' and ``S'' denote the big and the small black-hole solutions, respectively.
 At $T=T_{min}$ the free energy of the two solutions is the same.
 The free energy of the thermal gas phase is set to $F=0$.
 The direction along which the horizon location increases --
 which is represent by $\lambda_h$, the horizon value of the dilaton --
 is also shown in the figure.}
\label{fig2}
\end{figure}
The presence of the two types of black-holes is apparent from Figure \ref{fig1}(a),
which shows $T$ as a function of the horizon location.
It is also clear from the figure that there are no black hole solutions for $T<T_{min}$.
Although the small black-hole is always sub-dominant and has a negative specific heat,
its presence will turn out to be important to understand certain properties of $\etas$ in the following discussion.
Figure \ref{fig1}(b) shows the variation of the free-energy density $F$ as a function of the horizon radius.
Finally, by combining Figure \ref{fig1}(a) and Figure \ref{fig1}(b) one can parametrically solve for the free energy as a function of temperature.
The result is sketched in Figure \ref{fig2}, which also summarizes the phase structure of the system as the temperature is varied.

\subsubsection{Shear Viscosity of ihQCD}

\begin{figure}[h!]
 \begin{center}
\includegraphics[scale=1.2]{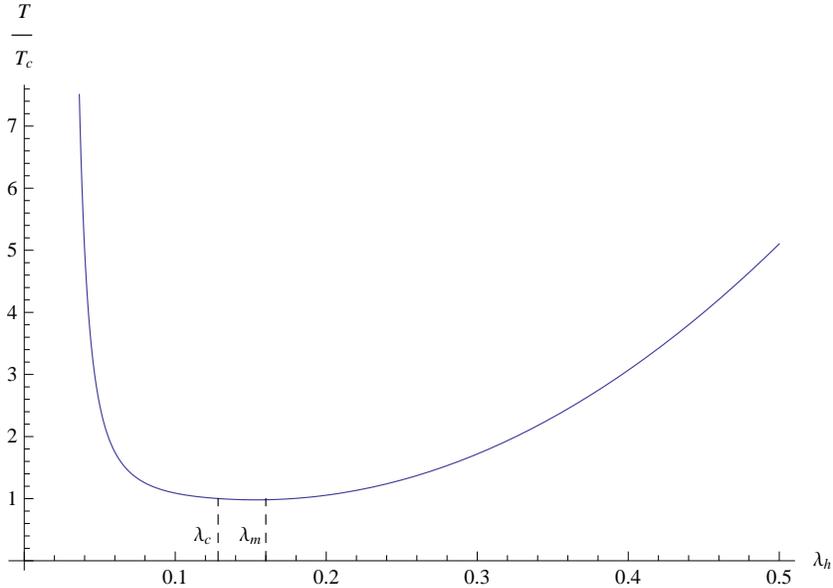}
 \end{center}
 \caption[]{Temperature as a function of $\l_h = e^{\sqrt{3/2}\Phi_h}$ in the ihQCD background.
 The labels $\l_c$ and $\l_m$ refer to the value of $\l_h$ at the critical and minimum temperature, respectively.}
\label{fig3}
\end{figure}

Given the general formula (\ref{etas}) and the dilaton potential (\ref{pot}), it is immediate to obtain $\etas$
as a function of the scalar at the horizon.
However, physically one is interested in having $\etas$ as a function of temperature, rather than $\Phi_h$.
Conversion from the latter to temperature is done by using either (\ref{Teq1}) or (\ref{Teq2}),
after solving Einstein's equations numerically for the background functions.
Figure \ref{fig3} shows the results of this calculation, for the choice of potential (\ref{pot}) which gives the best fit
to the available lattice data \cite{GKMN3}.
We note that the behavior of $T$ as a function of the horizon location is indeed of the form of Figure \ref{fig1}(a), with the minimum
separating the big black hole (on the left) from the small black hole branch (on the right).

Combining $T(\l_h)$ with the analytic expression (\ref{etas}) and the potential (\ref{pot}), one can now plot $\etas$
as a function of $T/T_c$, for a choice of the parameters $\beta$ and $\gamma$.
Unfortunately, we were unable to constrain the possible range of $\{\beta,\gamma\}$ with the available data, but instead chose
representative values (recall however that we want $\beta$ to be small enough so that the curvature corrections
in (\ref{fullaction}) remain perturbative).
Depending on the choice of the couplings $\{\beta,\gamma\}$, $\etas$ will then display different qualitative behaviors.

Two interesting fiducial cases --
which are representative of the behavior of the viscosity in a large portion of the phase space --
correspond to taking
$\b>0$, $0<\g<\sqrt{2/3}$ and $\b<0$, $\g<-\sqrt{8/3}$.
\begin{figure}[h!]
 \begin{center}
\includegraphics[scale=0.7]{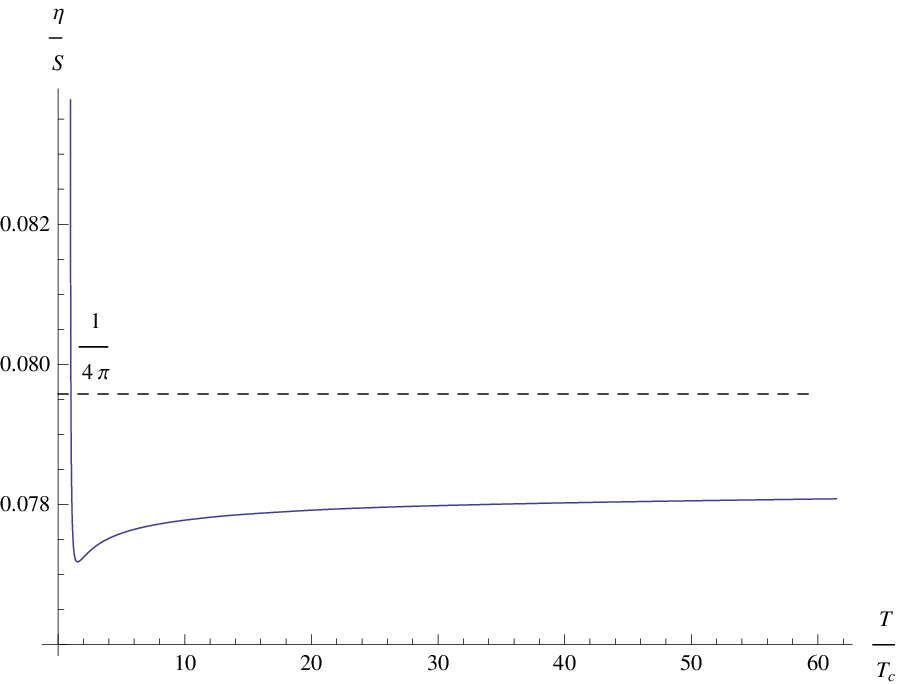}
\includegraphics[scale=0.7]{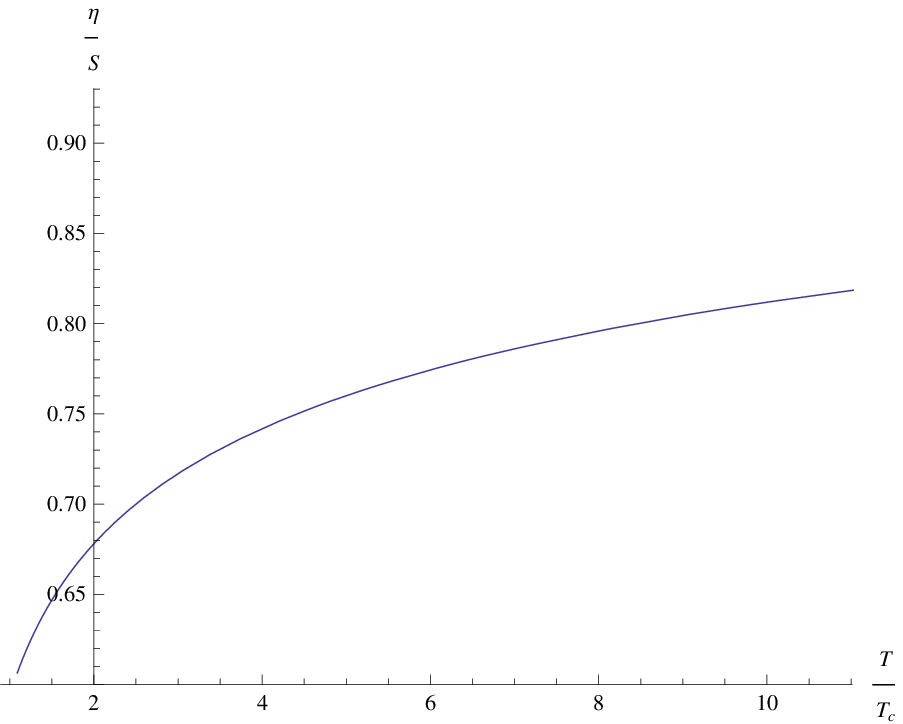}\\
\vspace{.5cm}
\hspace{.5cm}
(a)
\hspace{5cm} (b)
 \end{center}
 \caption[]{Plots of $\etas$ for theories with the scalar potential (\ref{pot}) and parameters chosen to
 give the best fit to lattice data \cite{GKMN3}. In (a) we have taken $\beta=0.1$ and $\gamma=\sqrt{3/2}$,
 while in (b) $\beta=-0.01$ and $\gamma=6/5\sqrt{3/2}$.
 Figure (a) shows that $\etas$ has a small minimum near $T \approx 1.8 T_c$.
 We emphasize that the size of the variation in $\etas$ can be reduced without changing its qualitative behavior, by taking $\beta$ sufficiently small.}
\label{ihqcdfig}
\end{figure}
In the first case, shown in Figure \ref{ihqcdfig}(a), $\etas$ displays a local minimum as a function of $T$ which, for the parameter
choices made there, appears around $T\approx 1.8 T_c$.
In the high temperature limit $\etas$ approaches the universal $\frac{1}{4\pi}$ value from below.
The behavior in the second case, depicted in Figure \ref{ihqcdfig}(b), is quite distinct.
There $\etas$ increases monotonically with temperature above $T_c$, increasing indefinitely as $T\to\infty$, just as in  perturbative QCD.
There is no local minimum appearing in the range $T>T_c$, and therefore $\etas$ acquires its minimum value \emph{at} $T=T_c$.
In both cases, we know from field theory studies of the hadronic phase that $\etas$ increases monotonically for $T<T_c$, as one probes
lower and lower temperatures. Thus, in Figure \ref{ihqcdfig}(b) we expect $T=T_c$ to correspond to the \emph{global}
minimum of the function $\etas(T)$.

In these two examples, the qualitative behavior of the viscosity is different not only near $T_c$, but also in the high-T regime.
We should note, however, that our holographic results can only be trusted up to a certain $T_{max}$,
above which the perturbative higher derivative expansion breaks down\footnote{A potentially useful way of characterizing
this break-down is by requiring $\ell^2\beta e^{\gamma\Phi}R \ll 1$, and evaluating this at the horizon.
This yields the relation $\frac{5}{3}\ell^2 \beta e^{\gamma\Phi_H} V(\Phi_H) \ll 1,$ from which one can compute $T_{max}$,
corresponding to the point at which the correction becomes $\mathcal O(1)$.
However, making this relation more precise is beyond the scope of this paper.}.
In particular, we won't be able to trust an arbitrarily large deviation from the universal $\frac{1}{4\pi}$ result.
Other interesting qualitative behaviors are also possible, in the remaining range of  $\{\beta,\gamma\}$.
For example, a mixture of the two aforementioned cases arises when $\b<0$ and $-\sqrt{8/3}<\g<0$.
In this case $\etas$ exhibits a local minimum above $T_c$ just like in the first case above, but its high-T behavior is the same as that of the second case.
We will provide a more detailed derivation of the different qualitative features of $\etas$ as a function of temperature in the next section,
where we will also discuss the boundaries of the various regions of phase space which lead to the distinct $\etas$ behaviors.

\subsection{Qualitative Features of the Shear Viscosity for Confining Backgrounds}

We would like to conclude this section by exploring some of the qualitative features of the temperature dependence of $\etas$
from a more general point of view.
For concreteness we will restrict our attention to dilaton potentials
that exhibit confining IR asymptotics (as $\Phi\to\infty$) as in the ihQCD case of the previous section,
\be\lab{IRasymp2}
V \to \frac{V_\infty}{\ell^2} \, e^{Q\Phi} \Phi^P +\cdots
\ee
and AdS asymptotics in the UV.
Furthermore, we will consider the following two possibilities for the behavior of the
potential in the UV:
\begin{enumerate}
\item[a.]
$V\to \frac{12}{\ell^2} + v e^{\, w \Phi} + \cdots,$ as $\Phi\to -\infty$
\item[b.]
$V\to \frac{12}{\ell^2} - \half m^2 \Phi^2 + \cdots $ as $\Phi\to 0$ \; .
\end{enumerate}
The first case (under the assumption that $w >0$ to ensure AdS boundary conditions) is precisely that of ihQCD-type backgrounds,
where the dilaton is massless and corresponds to a marginal deformation by the dimension-four operator $\tr F^2$.
The second case describes a massive dilaton of mass $m$, and corresponds to a deformation of the UV conformal theory
by an operator of scale dimension\footnote{Note that the normalization of the kinetic term for our scalar differs by a factor of four
from that of more standard AdS/CFT conventions, leading to a slightly different relation between the scalar mass and the
conformal dimension of the dual operator.}
\be\lab{delta}
\Delta = 2\le(1+ \sqrt{1+\frac{m^2\ell^2}{16}}\ri).
\ee
Thus, relevant deformations correspond to $m^2<0$, and the Breitenlohner-Freedman (BF) bound
is given by $m^2\ell^2=-16$.

In holographic constructions of the type we are considering, the flow of $\etas$ as a function of temperature
results from the way in which the near-horizon geometry changes as the horizon radius varies (it comes from sampling the \emph{phase space}
of the possible solutions to the theory).
In our setup, the horizon value $\Phi_h$ of the scalar field tracks the temperature of the system.
High temperatures then map to $\Phi_h\to-\infty$ in the a) type potentials, and $\Phi_h\to 0$ in the b) type potentials
we have just discussed, whereas `low' temperatures correspond to the region just above $T\sim T_c$.
We emphasize that the geometry which is relevant in this entire temperature range $T \gtrsim T_c$
is that of the big black-hole discussed in Section \ref{thermo}. Thus, the high-T behavior of $\etas$ will
correspond to the large horizon limit of the big black-hole geometry.
Below $T_c$, the thermal gas phase dominates, and we won't attempt to compute
$\etas$ directly in that regime (we will discuss the subtleties involved with such a calculation towards the end of this section).
Instead, we will extract information about the structure of $\etas$
-- and in particular any minima it has as a function of temperature --
by exploiting the existence of \emph{both} the big and small black-hole branches.
Thus, even though the small black-hole is not of direct physical interest -- it
is not thermodynamically favored -- it is still (indirectly) useful for probing $\etas$.

We start by working out the viscosity to entropy ratio in the regime which corresponds to the small black-hole branch.
In particular, we work in the $\Phi_h >> 1$ limit which describes an extremely small horizon radius (the far right of Figure \ref{fig3}).
In this regime, we can extract $\etas$ by plugging the IR potential (\ref{IRasymp2}) into our general formula (\ref{etas}), which gives
\be\lab{etasIR}
\frac{\eta}{s}   = \frac{1}{4\pi} + \b
\frac{V_\infty }{6\pi} \le(\frac{3\gamma Q}{4}-1\ri)  e^{(\gamma +Q)\Phi_h}\Phi_h^P+ \cdots \, ,
\ee
where we are omitting terms that are subleading in the large $\Phi$ limit.
Given that the small black-hole is not thermodynamically favored, we emphasize that
the only crucial piece of information we need to extract from this expression
is whether $\etas$ increases or decreases as $\Phi_h$ grows larger and the small black-hole shrinks.
We can see from (\ref{etasIR}) that as $\Phi_h$ grows,
the correction to $\etas$ tends to become\footnote{Analysis of the sub-leading terms shows
that it becomes large and positive even when $\g=4/3Q$.} {\em large and positive}
for $\{\b > 0, \g > 4/3Q\}$ and $\{\b < 0, \g < 4/3Q\}$,
while it tends into the {\em large and negative} direction in the opposite case (assuming in all these cases that $\gamma+Q>0$).
On the other hand, when $\g + Q < 0$ the sign of the exponential term changes, and
the correction term vanishes in the extreme small black hole limit, from above (below) when the product
$\b \left( \frac{3\g Q}{4} -1\right)$ is positive (negative).

We note that in all of this analysis $\b$ is assumed to be perturbatively small\footnote{
To ensure the absence of additional degrees of freedom associated with the higher derivative corrections in (\ref{fullaction}).}, $|\b |\ll 1$.
Of course when the higher derivative correction (\ref{etasIR}) becomes `too large,' the result can no longer be trusted -- for consistency one would have to include not only the four-derivative term in the action (\ref{fullaction}), but all other higher derivative corrections as well.
However, our expression for $\etas$ above will still be useful for the following two reasons.
First, in many phenomenological potentials of the type we are interested in, the leading IR behavior (\ref{IRasymp2})
doesn't necessarily set in at a very high value of $\Phi_h$.
Thus, in these cases there will be a region of validity for (\ref{etasIR}), which can be made
larger and larger by choosing $\beta$ (or $\gamma$) smaller.
Secondly, as it will become clear below, what we are really after is not the precise value of (\ref{etasIR}), but rather whether
$\etas$ tends upwards  or downwards on the small black-hole branch.

Next, we move on to discussing the high-temperature behavior of $\etas$ realized on the big black-hole branch\footnote{We are now probing
the near-horizon geometry of the big black-hole solution, in the large horizon limit.},
which maps to $\Phi_h \rightarrow -\infty$ and $\Phi_h \rightarrow 0$ for type a) and type b) potentials, respectively.
We analyze the two types of UV asymptotics for the potential separately.
In case a) one finds, for negative and large $\Phi_h$ (the black hole horizon is close to the runaway AdS region)
\be\lab{etasUVa1}
\frac{\eta}{s} \to \frac{1}{4\pi} \le[1 - 8\b e^{\g\Phi_h} + \cdots\ri]\, .
\ee
Thus, in the high-$T$ regime $\etas$ attains its universal $\frac{1}{4\pi}$ value, for $\g>0$.
Note that {\em it approaches this value from above (below) for $\b<0$ ($\b>0$)}. 
On the other hand, for $\g<0$, $\etas$ tends upwards (downwards) for $\b<0$ ($\b >0$).
We note that in the $\gamma<0$ case the result is only trustable up to a certain (negative) value of $\Phi_h$,
above which other higher order derivative corrections should also be taken into account. Thus, the expression for $\etas$
is only reliable up to a certain $T_{max}$.

For the b) type potentials, i.e. with a true AdS minimum at $\Phi=0$, one finds instead
\be\lab{etasUVb}
\frac{\eta}{s} \to \frac{1}{4\pi} \le(1 - 8\b\ri) -\frac{\b\g}{8\pi}\le(16+m^2\ell^2\ri)\Phi_h+ \cdots\, ,\qquad \Phi_h\to 0\, .
\ee
First of all, we learn that $\etas$ takes the value one expects\footnote{This conclusion is independent of the sign of $\g$,
because of the $\Phi_h \rightarrow 0$ limit.} from a curvature-squared correction (with no dilatonic scalar coupling)
in five dimensions \cite{Kats:2007mq}. Furthermore, recalling that the BF bound is $m^2\ell^2>-16$, we see that
{\em $\etas$ approaches its constant high-$T$ value from above (below) for $\b\g<0$ ($\b\g>0$)}.

There are a number of generic qualitative statements one can make about the running of $\etas$ with temperature
for these confining backgrounds:
\vspace{.3cm}

\paragraph{\bf Divergence of $\frac{d}{dT} \etas$ :}
First, it is known \cite{LongThermo} that for any potential that confines quarks at zero temperature,
there exists a temperature $T_{min}$ (below $T_c$) in the finite temperature theory where the small and big black-hole branches meet,
as shown in Figure \ref{fig1}a).
This implies that $\frac{d\Phi_h}{dT}$ diverges at this point and that $\etas$ is double-valued above $T_{min}$.
As a result, although $\etas$ itself is finite at $T_{min}$, by the chain rule, its derivative
 $\frac{d}{dT} \etas = \frac{d\Phi_h}{dT}\frac{d}{d\Phi_h} \etas$
will diverge there\footnote{One may worry that a diverging $\frac{d}{dT}\etas$ might
reflect a regime that can't be trusted. However, we can keep corrections to $\etas$ arbitrarily small by appropriately tuning
the perturbative coupling $\beta$. More importantly, the divergence of the derivative of $\etas$ at $T_{min}$ is a
result of the double-valued nature of $\Phi_h(T)$ -- the quantity $\frac{d}{d\Phi_h}\etas$ itself remains finite and small.
Thus, given that we are only interested in qualitative features of $\etas$ -- whether it increases or
decreases near $T_{min}$, and the sign of its slope there -- the divergence of $\frac{d}{dT}\etas$ is not expected to present a problem.}.
This is a \emph{generic} feature of $\etas$
that holds for any holographic background corresponding to a large N confining gauge theory at zero temperature.

This behavior is exemplified in Figures \ref{figdisc} and \ref{Cases3and4},
which show that the divergence occurs at the
meeting point of the big and the small black-hole branches. We should note that it is the thermal gas, and not one of the black-hole solutions, that is the thermodynamically preferred solution at this temperature and so the divergence in  $\frac{d}{dT} \etas$ does not actually correspond to a physical property of the shear viscosity. Even though it cannot be considered a physical characteristic of $\etas$, knowledge of the divergence of $\frac{d}{dT} \etas$
at $T_{min}$ will prove useful in analyzing the functional properties of $\etas.$ In particular, it will provide
{\em an holographic explanation for why $\etas$ assumes its minimum value at the transition point $T=T_c$
for confining gauge theories}, as we will see shortly.

There is also a particular case in which the divergence occurs on the border of a thermodynamically dominant branch and, as such,
{\it does} correspond to a physical property of $\etas$.
As described in \cite{LongThermo} and \cite{G1}, a second order confinement-deconfinement transition corresponds
to the case when $T_{min}$ and $T_c$ coincide. In this situation there is no small black-hole branch,
and the derivative of $\etas$ generically diverges precisely at the transition point $T=T_c$.
More precisely, $\frac{dT}{d\Phi}$
diverges at $T=T_c$ and therefore one generically expects that $\frac{d}{dT}\etas$ also diverges,
as long as $\frac{d}{d\Phi_h} \etas$ remains finite at that point.
This latter condition should be analyzed separately.

As a concrete example of this case, we take the generic backgrounds considered in \cite{G1}, which
exhibit a second (or higher) order Hawking-Page transition at a finite temperature $T_c$ at which $\Phi_h\to\infty$.
In order to find the behavior of $\etas$ around the transition point, one should first make sure that the perturbative approximation
in the higher derivative expansion is valid as $\Phi_h\to\infty$.
The scalar potential in the theories of \cite{G1} has the asymptotic behavior shown in (\ref{IRasymp2}) with
$P=0$ and $Q= \sqrt{8/3}$, together with exponentially decaying sub-leading terms as $\Phi\to\infty$,
\be\lab{contHP}
V(\Phi) \to e^{\sqrt{\frac83}\Phi} \le(1+ C\, e^{-\kappa \Phi}\ri), \qquad \kappa>0\, .
\ee
Here $C>0$ is a positive constant which does not play any role in the following discussion,
and the value of $\kappa$ determines the order of the continuous transition.

From the $\etas$ expression (\ref{etasIR}) we see that we can only make reliable statements about the behavior of the shear viscosity as
$T\to T_c$ ($\Phi_h\to\infty$) in the particular case  $\gamma<-Q=-\sqrt{8/3}$ (otherwise our perturbative approximation is
guaranteed to break down).
For this choice of parameters one then recovers the universal result $\etas \to \frac{1}{4\pi}$ as $T\to T_c$.
It is straightforward to work out the behavior of the {\em derivative} of $\etas$ by combining $\frac{d}{d\Phi_h}\etas$ and
$\frac{dT}{d\Phi}$.
The first can be calculated straightforwardly from (\ref{etasIR}).
The latter follows from the sub-leading analysis performed in \cite{G1}, which gives the following relation between temperature and $\Phi_h$,
\be\lab{contT}
T(\Phi_h) \to T_c + \tilde{C} e^{-\kappa \Phi_h}, \qquad \Phi_h\to\infty\, ,
\ee
where $\tilde{C}>0$.
Finally, combining these ingredients one obtains,
\be\lab{contT}
\frac{d}{dT} \frac{\eta}{s} \propto e^{(\gamma + \sqrt{\frac83} + \kappa)\Phi_h}, \qquad \Phi_h\to\infty\, ,
\ee
where the sign of the proportionality constant is positive (negative) for $\beta<0$ ($\beta>0$).
Thus we arrive at the conclusion that, although $\eta/s$ remains finite and reaches its universal $\frac{1}{4\pi}$ value
at the continuous transition,
its derivative $\frac{d}{dT}\etas$ diverges for  $-\sqrt{8/3}> \gamma > -\sqrt{8/3} -\kappa$, and it vanishes for $\gamma <  -\sqrt{8/3} -\kappa$.
We emphasize again that our perturbative approximation breaks down for $\gamma >   -\sqrt{8/3}$, in which case one cannot reliably
determine the behavior of $\etas$ at $T=T_c$.
This concludes our aside on the special case in which
the divergence occurs on the border of a thermodynamically dominant branch.
In the rest of the paper we will assume that $T_c \neq T_{min}$.
\vspace{.3cm}

\paragraph{\bf Conditions for local minima:}
Our general expression (\ref{etas}) makes the discussion of the presence or absence of a minimum for $\etas$ straightforward.
Local minima are defined by the following two conditions\footnote{Without loss of generality we assume that $dT/d\Phi_h$ is finite at these minima.},
\be\lab{mincond}
\frac{d}{d\Phi_h} \etas =0 \, , \qquad \frac{d^2}{d\Phi_h^2} \etas >0 \, ,
\ee
which, when re-instating the generic higher derivative coupling $G(\Phi)$, can
be translated into analyticity conditions on $G(\Phi)$ and the dilaton potential $V(\Phi)$ using (\ref{etas}).
In particular, extrema of $\etas$ are determined by the locii $\Phi_h$ at which
\be\lab{extrema}
\left(\frac34 G^{\, \prime} (\Phi) V^{\, \prime}(\Phi_h) - G(\Phi) V(\Phi)\right)'\Big|_{\Phi=\Phi_h}=0 \, , \,
\ee
an expression which is valid for arbitrary potentials (which allow for AdS asymptotics) and scalar couplings $G$, and not just the
particular setup of this section.
In terms of $G(\Phi) = e^{\gamma\Phi}$, the condition becomes
\be
\frac34 \g V''(\Phi_h) + (\frac34 \g^2-1)V^{\, \prime}(\Phi_h) - \g V(\Phi_h) = 0 \, .
\ee
Thus, in order for $\etas$ to have an extremum at a certain temperature, a necessary condition\footnote{This is also
sufficient provided that $dT/d\Phi_h$ is finite at this point.} is that the dilaton potential $V(\Phi)$
possesses at least one solution to equation (\ref{extrema}).
Clearly, for such a solution to correspond to a minimum one should further ensure that the second derivative is positive there.
These two conditions provide a simple criterion for the existence of extrema, given an arbitrary scalar field profile in the class
of theories (\ref{fullaction}).
\vspace{.3cm}

\paragraph{ \bf Presence of a global minimum:}
The discussion in the previous paragraph concerns general conditions for the existence of local minima.
However, we can also ask what are the conditions for the existence of a \emph{global} minimum, given a
confining potential with the IR and UV asymptotics described above.
To answer this question, it turns out to be useful to recall the behavior of $\etas$
in the two opposite limits of an extremely small and extremely large black-hole.
We will restrict our attention to the case in which the function $\etas(T)$ is monotonic on the small black-hole branch,
an assumption which is satisfied in all of the holographic backgrounds we consider in this paper.
Our analysis will also make use of the fact that the derivative of $\etas$ approaches $\pm \infty$
as $T \rightarrow T_{min}$ along the two black-hole branches, as we discussed above.
We will divide the discussion into two cases:
\begin{enumerate}
\item{\bf Discontinuous case:}
Here we discuss the situation in which $\etas$ displays a global minimum appearing \emph{exactly at} $T=T_c$.
If $\etas$ is a {\em monotonically increasing} function of temperature on the entire big black-hole branch,
reaching its maximum value as $T\to\infty$ from below,
then clearly it will reach a minimum value (on the big black-hole branch) at $T=T_c$, where the big black hole
solution ceases to be thermodynamically favored.
To work out the behavior of $\etas$ in the range $T<T_c$ one would then need to switch to the thermal gas solution.
Calculating $\etas$ directly in this regime is technically very challenging.
However, as outlined in the introduction, there are qualitative field theory arguments which tell us that $\etas$ should
start increasing again below $T_c$, as the temperature continues to drop.
Thus, we conclude that in this case $T=T_c$ corresponds to a global minimum for $\etas$.
This behavior is shown schematically in Figure \ref{thermal}, from which it is also clear that the derivative of $\etas$
is \emph{discontinuous} at $T_c$.

\begin{figure}
 \begin{center}
\includegraphics[scale=0.4]{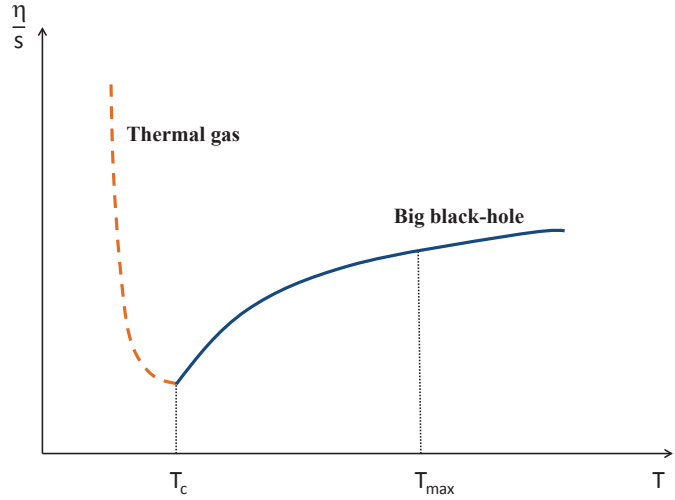}\\
 \end{center}
 \caption[]{A cartoon of the behavior of $\etas$ in the big black-hole phase (to the right of $T_c$) and
 in the thermal gas phase (for $T<T_c$) for the case where the big black-hole phase is monotonically increasing above $T_c$. We denote by $T_{max}$ the temperature above which the correction to $\etas$ is no longer perturbative,
 and our approximation breaks down.}
\label{thermal}
\end{figure}

\begin{figure}
 \begin{center}
\includegraphics[scale=0.4]{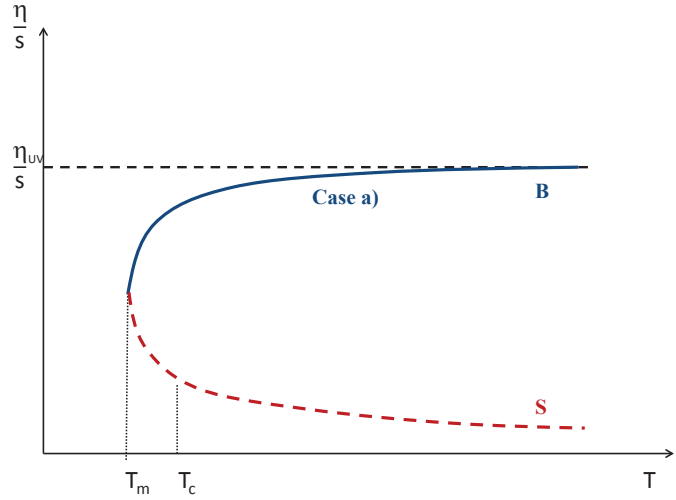}\\
(a)\\
\vspace{.5cm}
\includegraphics[scale=0.4]{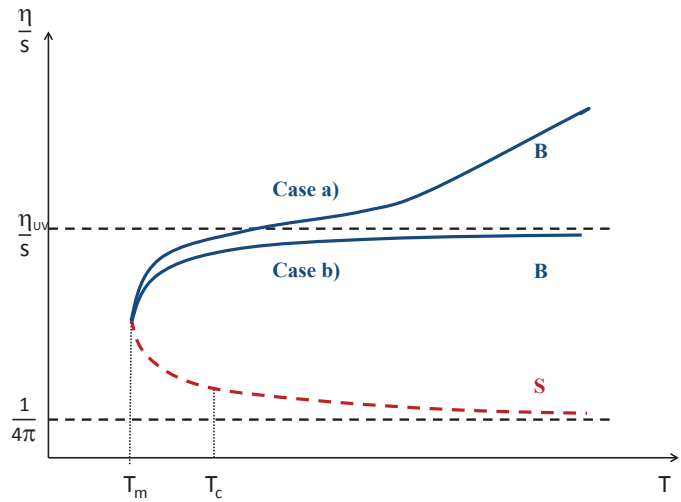}\\
(b)
 \end{center}
 \caption[]{(a) A cartoon of the case when (\ref{casei}) is satisfied, for the type a) potential.
 (b) A cartoon of the case when (\ref{caseii}) is satisfied, for the type a) and type b) potentials
 labeled respectively by ``case a'' and ``case b.''
 In both figures $T_m$ denotes the temperature where the small (S)
 and the big (B) black-hole branches meet and $T_c$ denotes the temperature of the confinement-deconfinement transition.
 The high temperature value of the viscosity to entropy ratio is denoted by $\frac{\eta_{UV}}{s}$.}
\label{figdisc}
\end{figure}

Graphically it is clear that, in order for this to happen, on the big black-hole branch
$\etas$ must approach its high-$T$ value from below,
and furthermore its derivative on this branch must approach $\frac{d}{dT} \etas \to +\infty$
at $T_{min}$, where the big and the small black-hole branches meet.
Given our assumption that $\etas$ is a monotonic function of $T$ on the small black-hole branch,
in order to satisfy the second criterion it suffices that $\etas$ decreases as $\Phi_h$ gets larger
(in the extreme $\Phi_h\gg 1$ limit) on the small black-hole branch.
By examining  (\ref{etasIR}) and (\ref{etasUVa1}) we see that this requires\footnote{Note that we need
the derivative of $\etas$ to be positive definite in the asymptotic high-$T$ region (on the big black-hole branch).
Since $dT/d\Phi_h$ is negative definite on the big black-hole branch,
we see from (\ref{etasUVa1}) that this requires $\b\g>0$.}
either of the following two cases:
\bea
\lab{casei}
[i.]\,\,\, \b&>&0, \qquad 0<\g<\frac{4}{3Q}\, ,\\
\lab{caseii}
[ii.]\,\,\, \b&<&0, \qquad \g<-Q\, .
\eea
We plot these two cases schematically in Figure \ref{figdisc}(a) and \ref{figdisc}(b).
Figure \ref{figdisc}(a) shows the behavior of $\etas$ on the big (B) and small (S) black-hole branches for a potential
whose UV asymptotics are described by case a). In Figure \ref{figdisc}(b) on the other hand we include both cases a) and b).

\begin{figure}
 \begin{center}
\includegraphics[scale=0.4]{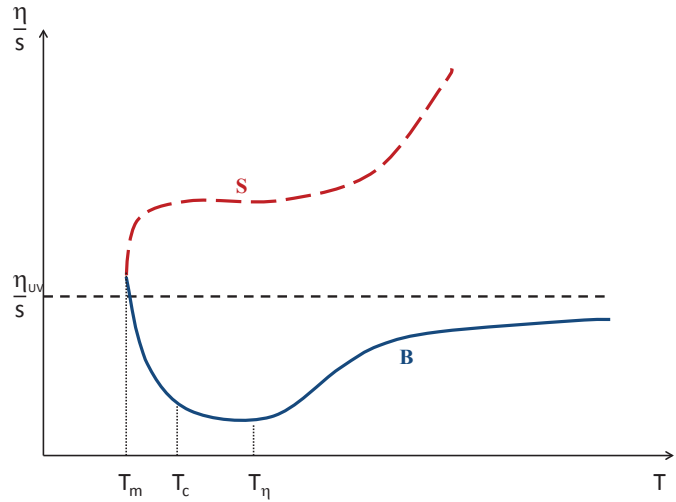}\\
(a)\\
\vspace{1cm}
\includegraphics[scale=0.4]{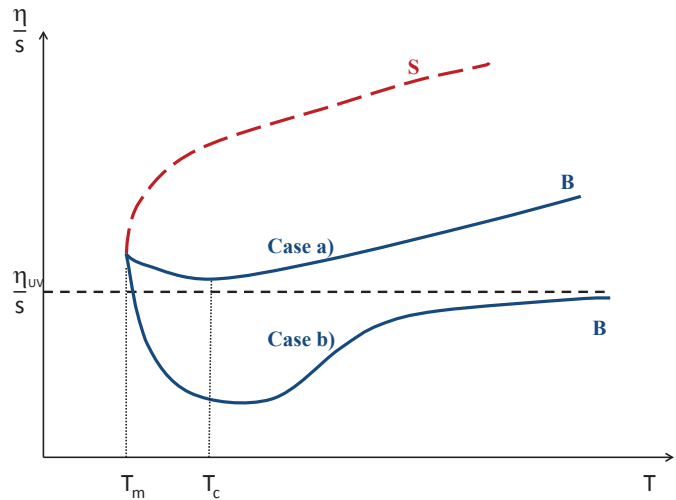}\\
(b)
 \end{center}
 \caption[]{(a) A cartoon of the case when (\ref{caseiii}) is satisfied. (b) A cartoon of the case when (\ref{caseiv}) is satisfied.
 $\frac{\eta_{UV}}{s}$ denotes the high-$T$ value of $\etas$, $T_m$ denotes the minimum temperature where the small (S)
 and the big (B) black-hole branches meet and $T_c$ denotes the temperature of the confinement-deconfinement transition.
 In Figure (b) we plot both the type a) and type b) potentials labelled by ``case a'' and ``case b'' respectively.}
\label{Cases3and4}
\end{figure}

\item{\bf Continuous case:}
Here the $\etas$ curve has a global minimum on the big black-hole branch at some temperature $T_\eta$
satisfying $T_\eta >T_c$.
Note that at that point, $\etas$ and all of its derivatives are \emph{continuous}.
For this to happen, clearly $\etas$ must increase as the temperature is raised above $T_\eta$,
approaching its high-T value \emph{from below}.
More importantly, $\etas$ must also increase below $T_\eta$, as the temperature gets closer and closer to $T_{min}$.
The latter condition translates into the requirement that the derivative on the big black hole branch obeys
$\frac{d}{dT}\etas
\to -\infty$ at $T_{min}$.
Finally, it suffices that $\etas$ tends upwards on the small black-hole branch\footnote{Recall that we are assuming that $\etas$ is monotonic there.},
in the $\Phi_h\gg 1$ limit.
Investigation of (\ref{etasIR}) and (\ref{etasUVa1}) reveals that one needs
\bea
\lab{caseiii}
[i.]\,\,\, \b&>&0, \qquad \g>\frac{4}{3Q}\, ,\\
\lab{caseiv}
[ii.]\,\,\, \b&<&0, \qquad -Q<\g<0\, .
\eea
We provide schematic drawings of these two cases in Figure \ref{Cases3and4}(a) and \ref{Cases3and4}(b).
Note also that (\ref{caseiii}) corresponds to the case plotted in Figure \ref{ihqcdfig}(a).
Of course one should also incorporate the fact that, in confining theories, the thermodynamically preferred background changes below $T_c$.
Thus, in order to be able to observe a global minimum located at some temperature $T_\eta$,
one should ensure that $T_\eta > T_c$, as shown in Figure \ref{Cases3and4}(a).
\end{enumerate}
The four cases depicted in Figures \ref{figdisc}(a), \ref{figdisc}(b), \ref{Cases3and4}(a) and \ref{Cases3and4}(b)
cover most of the physically interesting situations. Qualitatively distinct behavior (for example a maximum rather than a minimum)
for $\etas$ arises in the other ranges of $\b$ and $\g$, but we will not discuss it here.
Finally, we should note that allowing for a more general scalar coupling $G(\Phi)$ to the higher derivative term (\emph{i.e.} of the
racetrack-type) would modify -- and potentially significantly complicate -- the analysis of the existence of a global minimum.
We will not attempt this here.

\subsubsection{The Thermal Gas Phase}
We would like to close this section by discussing briefly
the main difficulties involved with calculating $\etas$ on the thermal gas background which describes temperatures below the deconfinement
transition.
We emphasize again that in our analysis we did not explicitly compute $\etas$ below $T_c$, but rather used knowledge of
its behavior in the hadronic phase.
The main difficulties involved with a direct computation for $T < T_c$ stem mostly from the need to consider $1/N^2$ corrections.

To see why this is so, note that the thermal gas solution
has no horizon, hence the $\cO(N^2)$ contribution to the entropy that would come from evaluating the action on
the classical saddle\footnote{Note the identification $1/G_5 \propto N^2$ in front of the action.
Thus, whenever the on-shell action on a classical saddle is non-vanishing, automatically all of the thermal
functions are proportional to $N^2$.}
vanishes.
However, the finite temperature in the background
generates thermal fluctuations of the graviton gas.
Thus, the entropy in this case should be calculated by computing the determinant of fluctuations around the classical thermal gas saddle,
and as such it is $1/N^2$ suppressed with respect to the $\cO(N^2)$ black-hole contribution.
As a result, the entropy of the thermal gas is $s_{tg} = \cO(1)$.

Now, let's move on to the calculation of $\eta$ on the thermal gas background, which can be done by
considering fluctuations of the transverse traceless gravitons -- in particular, $\etas$ can be related
to the flux of transverse gravitons through the horizon, which is proportional to $e^{3A_h} N^2$.
On the other hand, the {\em small} black-hole background  asymptotes to the (Lorentzian) thermal gas in the limit of vanishing horizon radius.
This tells us that the leading $\cO(N^2)$ contribution to $\eta$ also vanishes, in the same way in which the $\cO(N^2)$
contribution to the entropy of the thermal gas vanished.
In order to find the finite contribution one has to consider Witten diagrams with quantum loops,
which lead to $\eta_{tg}\propto \cO(1)$.
Thus, although $\etas$ in the thermal gas is of the same order as that for any black-hole background,
in order to calculate it one needs to compute the $1/N^2$ contributions to both $\eta$ and $s$.
Such a calculation would involve computing one-loop diagrams in supergravity.
This would naively require knowing the spectrum of all supergravity fields which could run in the loops\footnote{Note, however,
that one-loop contributions to hydrodynamic long-time tails do not require a detailed understanding of the spectrum of
supergravity fields \cite{CaronHuot:2009iq}. Perhaps a similar simplification occurs for one-loop contributions to $\eta/s$.}
and as such, knowing a genuine string theory embedding of any particular model.
This one-loop calculation appears to be a difficult task and we don't attempt it here.

\section{Discussion}
\label{Conclusions}

Before summarizing our results for the temperature dependence of the shear viscosity in holographic models, we would first like to discuss several properties of $\eta/s$ in the QCD plasma. Although the viscosity to entropy ratio in the QGP phase is expected to be very small and roughly
comparable to $\frac{1}{4\pi}$, its precise value is still largely uncertain.
The collective flow observed at RHIC and LHC is analyzed by performing a Fourier decomposition of
the particles' angular distribution, whose Fourier coefficients $v_n$ are sensitive to transverse momentum $p_T$.
Restrictions on the possible range of $\etas$ are then extracted from the data either by
comparing the momentum-dependent elliptic flow coefficient $v_2(p_T)$ with results obtained from viscous
hydrodynamical calculations\footnote{More precisely, one needs a `hybrid' code which combines viscous hydrodynamics of the QGP phase
with a realistic model of the late hadronic stage (see e.g. \cite{Heinz:2011kt}).},
or by fitting the centrality dependence of the average $p_T$ integrated elliptic flow.
Both approaches yield bounds on $\etas$ which are somewhat close to the universal $\frac{1}{4\pi}$ value,
with the analysis of \cite{Song:2010mg} giving $ 4 \pi \, \etas \leq 2.5$.\\

\noindent {\bf Uncertainties of $\etas$}\\
A number of challenges, both theoretical and experimental, have to be overcome in order to determine $\etas$ more precisely.
Uncertainties on the experimental side come from non-flow correlations introduced, for example, by jets and resonance decays. 
On the theoretical side, uncertainties are due to various assumptions on the choice of initial conditions, initial state
pressure gradients and event-by-event fluctuations.
As an example, it was realized only recently that the assumption that higher order harmonics
were negligible was a poor one -- the shape of the fireball can fluctuate from one event to the next, even at a fixed impact parameter.
The resulting irregular pressure gradients are \emph{not} symmetric with respect to the reaction plane,
and can in fact induce higher harmonic flow patterns.
The elliptic flow direction and magnitude can also fluctuate event-by-event.
Interestingly, the measurement of higher harmonics can play an important role in reducing the degeneracy between the shear viscosity and
initial conditions, and appears to give tighter limits on $\etas$.
In fact,  a combined analysis \cite{Qiu:2011hf} of elliptic and triangular flow coefficients ($v_2$ and $v_3$, respectively)
in Pb $+$ Pb collisions at LHC was shown to favor a small shear viscosity $\etas \simeq 0.08$,  disfavoring the considerably larger
value of $ \etas \sim 0.2$ (we refer the reader to \cite{Muller:2012zq} for a more detailed discussion of these issues).
Although none of the models currently used describes perfectly all experimental flow data,
the expectation is that -- with the new high precision data coming from LHC -- it will be possible
to cut down on such uncertainties and arrive at a precision measurement of the transport properties of the QGP.\\

\noindent {\bf The temperature dependence of $\etas$}\\
Another important issue for disentangling the physics of the QGP is that of the possible temperature dependence of $\etas$,
which is the main focus of our analysis.
Although most hydrodynamical simulations of the QGP assume a constant value for $\etas$,
a number of studies -- mostly qualitative in nature --
have begun to examine the possible relevance of temperature in LHC and RHIC heavy ion collisions
\cite{Niemi:2011ix,Nagle:2011uz,Shen:2011kn,Bluhm:2010qf}.
First results for the shear and bulk viscosity obtained via lattice QCD simulations
were put forth in \cite{Nakamura:2004sy,Meyer:2007ic}, where it was shown that $\etas$
remains close to its universal $\frac{1}{4\pi}$ value for a range of temperatures not far above $T_c$.
On the other hand, kinetic theory results for QCD at \emph{weak coupling} \cite{Arnold:2000dr,Arnold:2003zc}
imply $\etas >1$, for the QCD running coupling constant in the range $\alpha_s \lesssim 0.25$.
Elliptic flow values at LHC energies may be sensitive to the temperature behavior of $\etas$
in the QGP phase, although insensitive to it in the hadronic phase \cite{Niemi:2011ix}.
Moreover, it was argued in \cite{Csernai:2006zz} that $\etas$ should have a minimum at (or near) the QCD phase transition.
This is expected because $\etas$ increases with decreasing temperature in the hadronic phase \cite{Gavin:1985ph,Prakash:1993bt}, while asymptotic
freedom dictates that it increases with temperature in the deconfined phase -- thus leading to at least one minima somewhere in the intermediate region.
An interesting study of shear and bulk viscosities for a finite temperature, pure gluon plasma was performed in \cite{Bluhm:2010qf},
where the authors used a phenomenological, quasi-particle model based on an effective kinetic theory description.
At large temperatures, the results of \cite{Bluhm:2010qf} reproduce parametrically the transport behavior
observed in perturbative QCD calculations.
As the non-perturbative regime is approached, their analysis is in agreement with lattice QCD results,
with a decrease of $\etas$ with temperature and a minimum  at $T \gtrsim T_c$.
Independently of the assumptions behind particular choices of model and simulation schemes, it is clear
that a more systematic understanding of the temperature dependence of $\etas$
is an important ingredient
towards a better descriptions of the dynamics of the QGP.

\subsection{Summary of Results}

The aim of this paper was to initiate a systematic study of the temperature dependence of the shear viscosity in the strongly coupled
QGP (and gauge theory plasma more generally), in the framework of the holographic gauge/gravity duality.
We were largely motivated by the potential sensitivity of elliptic flow measurements to thermal variations of $\etas$,
but were also interested in gaining further insight into the geometric interpretation of a non-trivial temperature flow
for the shear viscosity.
We restricted our attention to theories of gravity coupled to a scalar field in the presence of higher derivative corrections,
\beq
\label{lagrangian}
\mathcal{L} = R - 2 (\nabla\Phi)^2 + V(\Phi)
+
\beta \, G(\Phi)\, R_{\mu\nu\rho\sigma}R^{\mu\nu\rho\sigma} \, ,
\eq
under the assumption that the coupling $\beta$ is perturbatively small\footnote{From now on we will set
the AdS length scale $\ell$ equal to one.}.
Higher order curvature corrections
are well-known to push $\etas$ away from its universal $\frac{1}{4\pi}$ value.
Moreover, once they are coupled to a non-trivial scalar field profile, they lead generically to a
temperature flow for $\etas$, as already seen in \cite{Buchel:2010wf} and emphasized in \cite{Cremonini:2011ej}.

The correction to $\etas$ -- which is parametrized entirely in terms of horizon data --
can be expressed in terms of the potential and the horizon value $\Phi_h$ of the scalar field,
and for (\ref{lagrangian}) takes the relatively simple form (see Section II for details of the derivation)
\be
\etas
= \frac{1}{4\pi} \left[1+\frac23
\beta \left(-G(\f_h)V(\f_h)+ \frac34 G^{\, \prime}(\f_h)V^{\, \prime}(\f_h)\right)  \right]\, .
\label{etasgeneral}
\ee
Note that thanks to this expression, determining the presence of \emph{local} minima for $\etas$ is entirely straightforward.
In particular, extrema are found by minimizing (\ref{etasgeneral}) with respect to\footnote{Assuming that $dT/d\Phi_h$ is finite.} $\Phi_h$,
and therefore correspond to the zeros of the relation (\ref{extrema}).
We emphasize that such a condition is rather generic,
given the broad assumptions behind (\ref{etasgeneral}) -- essentially the requirement of AdS asymptotics.

Armed with the general expression (\ref{etasgeneral}), we have analyzed a number of holographic setups.
For the majority of our analysis, we have restricted our attention to the special case of
$G(\Phi) = e^{\gamma\Phi}$, for which the viscosity to entropy ratio reduces to
\be
\etas
= \frac{1}{4\pi} \left[1+\frac23 \,
\beta \left(-V(\f_h)+ \frac34 \, \gamma \, V^\prime(\f_h)\right)  e^{\gamma \Phi_h} \right]\, .
\lab{etasconcl}
\ee
In this class of theories, the (horizon) quantity $\Phi_h$ tracks the temperature of the system.
As a result, the behavior of $\etas$ as a function of temperature is controlled by the scalar field coupling
to the curvature corrections -- parametrized here by $\gamma$ and $\beta$ -- as well as by the specific functional form of the potential.

As a simple consistency check of our result, we note that for the case of a non-dynamical scalar our expression
(\ref{etasconcl}) reduces to a constant and reproduces the standard result for an AdS black-brane in pure gravity with curvature-squared terms.
On the other hand, an analytically tractable example which gives rise to non-trivial temperature dependence
is that of a single exponential potential,  $V \propto e^{\, \alpha \Phi}$.
Black-brane solutions in this system have a temperature of the form
$T \propto e^{- \frac{4}{3\alpha} \left(1-\frac{3\alpha^2}{8}  \right) \Phi_h}$.
With this choice of potential,
our expression (\ref{etasconcl}) implies a monotonic temperature flow for $\etas$, whose precise structure
is shown in (\ref{etaSCR}) and is dictated by the range and signs of $\beta$ and $\gamma$.
Interestingly, when $\gamma = - \alpha$ the viscosity to entropy ratio takes on a constant value again, with no temperature dependence.
In particular, for the specific choice $\gamma = - \alpha = \sqrt{2/3}$ the model can be obtained via a $U(1)$
dimensional reduction of a six-dimensional theory of pure gravity with a negative cosmological constant,
in the presence of $R^2$ corrections. As expected, our correction to $\etas$ in this case reduces to that for a pure $AdS_6$ black brane,
providing a rather non-trivial check of our result (\ref{etasconcl}).
The absence of temperature dependence in this case is then a simple consequence of the fact that the five-dimensional theory
inherits the scale invariance of the parent six-dimensional one (see also the discussion in \cite{Blaise}).

The behavior of $\etas$ becomes more interesting -- and more relevant to the physics of the QGP --
for theories which undergo a confinement--deconfinement transition at some critical temperature $T_c$.
In our analysis we have worked mostly with potentials constructed `phenomenologically' by requiring their
IR and UV asymptotics to match with expectations from QCD,
as in the case of the \emph{Improved Holographic QCD} (ihQCD) model discussed in Section \ref{ihqcd}.
However, we emphasize that a number of our results hold more broadly, and can be easily generalized.
Plots of the temperature dependence of $\etas$ in the ihQCD model for fiducial values of $\{\beta,\gamma\}$ are shown in Figure \ref{ihqcdfig}.
On the left, we note the presence of a small minimum slightly above $T_c$, above which $\etas$ approaches $\frac{1}{4\pi}$.
The behavior for $T<T_c$ was not computed directly, but rather extrapolated using the fact that
$\etas$ is expected to increase in the hadronic phase, as the temperature is lowered.
The figure on the right, on the other hand, exhibits a minimum exactly at $T=T_c$, with $\etas$ growing indefinitely as the temperature is raised.
A cartoon of the same type of behavior -- expected to describe more general settings -- is shown in Figure 6.

For theories which confine quarks at zero temperature, the thermodynamic structure of the corresponding
finite-temperature theory plays an interesting role in determining the behavior of $\etas$,
and enables one to make generic \emph{qualitative} statements about its temperature flow.
In this class of models, the geometry which is favored thermodynamically for $T>T_c$
is that of a \emph{big} black hole, while below $T_c$ the dominant phase corresponds to a thermal gas.
However, a third solution appears for $T>T_{min}$ (with $T_{min} < T_c$), which describes a \emph{small} black-hole solution.
For the ihQCD model, the existence of the two black-hole branches can be seen in Figure \ref{fig3},
with the minimum separating the big black hole (on the left) from the small one (on the right).
Although the small black-hole is never thermodynamically favored -- and is therefore not of any direct physical relevance --
it can still be useful for probing $\etas$ in the physically relevant regime $T>T_c$.
In fact, the coexistence of the small and big black-hole solutions for $T>T_{min}$ implies that $\etas$
is double-valued there, and that its derivative
diverges at the point $T=T_{min}$ where the two branches meet.
These features can be seen for example in Figure 7(a).
By examining the rough dependence of $\etas$ on temperature on the two branches
and the sign of $\frac{d}{dT} \etas$ as $T_{min}$ is approached, one can then reconstruct
whether $\etas$ exhibits a minimum in the $T \gtrsim T_c$ range.

Following this strategy, in Section IIIC we have derived a set of \emph{geometric}
conditions for the existence of a global minimum for $\etas$ in this class of models\footnote{Under the (mild) assumption that $\etas$
is monotonic on the small black-hole branch.}.
We have chosen to work with dilatonic potentials which have confining IR asymptotics and exhibit two distinct behaviors in the UV,
both consistent with AdS asymptotics
-- the type a) and type b) potentials corresponding, respectively, to a massless and massive scalar.
The requirement -- from asymptotic freedom -- that $\etas$ approaches its high-T value from below
fixes the signs and ranges of the higher derivative couplings $\beta$ and $\gamma$.
Combined with the requirement that the zero-temperature theory is confining,
it \emph{guarantees} the existence of a global minimum for $\etas$ at or above the critical temperature,
in a certain portion of the phase space. Furthermore, whether the (global) minimum
is located exactly at $T_c$ or above it can be determined by
looking at the sign of the divergence of $\frac{d}{dT} \etas$ at $T_{min}$, the point where the big and small black hole branches meet.

In Section IIIC we have provided a geometric classification of these two cases, which we emphasize are physically distinct.
The parameter space of the `\emph{discontinuous}' case, describing a global minimum exactly at $T_c$,
is summarized in (\ref{casei}) and (\ref{caseii}), and shown schematically in Figure 7.
The `\emph{continuous}' case, with a global minimum at some temperature $T > T_c$, is
described instead in (\ref{caseiii}) and (\ref{caseiv}), and shown in Figure 8.
We note that both cases have $\beta \gamma >0$ (for the precise range of $\gamma$ we refer the reader to Section IIIC).
The two types of UV asymptotics, cases a) and b), are also shown in the figures. The high-temperature behavior of $\etas$,
and in particular whether it approaches a constant value or increases indefinitely with temperature, is determined
by the values and ranges of the $\beta$, $\gamma$ parameters.
Another generic result of our analysis is that when the confinement-deconfinement phase transition is {\em second (or higher) order}
the derivative $\frac{d}{dT} \etas$ diverges or vanishes at $T_c$, depending on the value of the parameter $\gamma$,
while $\eta/s$ remains finite and approaches its universal $\frac{1}{4\pi}$ value\footnote{This result
is valid as long as the perturbative approximation can be trusted. See section IIIC for details.}.

One should note, however, that the holographic demonstration of the existence of a global minimum in confining backgrounds is incomplete.
In particular, it is desirable to find an independent holographic reason for the requirement that $\beta\gamma$ should be positive,
needed for the presence of the global minimum in the continuous/discontinous cases discussed above.
In fact, although we always insisted on keeping the correction to $\etas$ perturbatively small (by appropriately turning down $\beta$),
we relied on the fact that $\frac{d}{dT}\etas$ diverges near $T_{min}$. Thus, one may worry that any conclusion drawn from
using $\frac{d}{dT}\etas$ in a regime where it is very large might not be valid.
While this is a fair concern, we should note that it is $\frac{d\Phi_h}{dT}$ which diverges, with $\frac{d}{d\Phi_h}\etas$
remaining finite (and small).
For completeness, it would also be desirable to perform the analogous $\etas$ calculation in the hadronic phase ($T<T_c$), and
derive holographically the monotonically decreasing behavior of $\etas$ expected from field theory.
Finally, we should caution the reader that a more generic higher derivative scalar coupling $G(\Phi)$ would
give rise to a more complicated structure for $\etas$ (\emph{i.e.} imagine allowing for racetrack-type terms) and
might invalidate some of our arguments for the presence of a global minimum. Clearly, our analysis would have to be re-evaluated
in such situations.
Regardless, we would like to emphasize that -- for a certain class of potentials, and for simple
higher derivative couplings of the form $ G \propto e^{\gamma\Phi}$ -- we have translated the issue
of the existence of a minimum for $\etas$ into a geometric one, and identified the holographic conditions which would
guarantee its presence, complementing the field-theoretic arguments of \cite{Kovtun:2011np}.

In order for our analysis to be of more direct relevance to studies of the QGP, it is crucial to find ways to place
restrictions on the allowed ranges of the couplings $\beta$ and $\gamma$, whether with theory or experiment.
This would cut down on the large phase space for the temperature behavior of $\etas$, and eliminate some of the model dependence inherent
in holographic setups of this type -- in particular when higher derivative terms are present.
At the moment, however, this is very challenging.
In the future, it might be feasible to constrain models by examining the effects of higher derivative corrections
on the remaining transport coefficients (including those of second order hydrodynamics), and then
simultaneously fitting to the available data.
While this should be possible in principle, at the moment it presents a serious challenge --
the bulk viscosity flows with temperature already without the need for higher derivative terms,
and very little is currently known from holography about second order transport coefficients in the presence of higher derivative corrections.
On the other hand, it may be possible to restrict the values of the parameters with more phenomenological considerations.
As an example, at least in principle we can fit the UV behavior in the type a) potentials we discussed
to that predicted by perturbative QCD, which would fix the values of both couplings $\beta$ and $\gamma$.
However, we should keep in mind that our analysis is not valid to arbitrarily high temperatures (it breaks down when the higher
derivative interactions are no longer perturbative), and therefore any direct comparison with UV physics should be taken with a grain
of salt.
We leave these questions open for future work.

\section*{Acknowledgments}

We thank Allan Adams, Peter Arnold, Nabamita Banerjee, Alex Buchel, Anatoly Dymarsky, Suvankar Dutta, Blaise Gouteraux, Elias Kiritsis,
Jamie Nagle, Paul Romatschke, Dam Thanh Son, Andrei Starinets, Amos Yarom and Urs Wiedemann for interesting discussions.
S.C. is grateful to the Newton Institute for hospitality during the
Mathematics and Applications of Branes in String and M-theory workshop
while this work was underway.
The work of S.C. has been supported by the Cambridge-Mitchell Collaboration in Theoretical
Cosmology, and the Mitchell Family Foundation.
U.G. is grateful to the University of Crete, the organizers of the "Cosmology and Complexity 2012" workshop in Hydra, Greece,
and the IMBM (Istanbul Center for Mathematical Sciences) where parts of this work have been completed.
The work of P.S. has been supported by the U.S. Department of Energy under Grant No. DE-FG02-97ER41027.
This research was  supported in part by the National Science Foundation under Grant No. PHY05-51164.

\appendix

\section{The Method of Phase Variables}
\label{PhaseVars}

For the study of dilatonic black hole solutions, it turns out to be particularly
convenient to adopt the \emph{phase variables} method developed in \cite{LongThermo}, which we briefly review here.
With the black brane ansatz
\begin{equation}
\label{BHu}
  ds^2 = f^{-1}(r) \, dr^2 + e^{2A(r)}\le( d\vec{x}^2 - f(r) \, dt^2  \ri), \qquad \Phi =   \Phi(r) \, ,
\end{equation}
the Einstein and dilaton equations of motion that follow from the action (\ref{action0})
can be put in the following simple form of five {\em first-order} equations,
\bea\label{Ap}
\frac{dA}{dr} &=& -\frac{1}{\ell} \, e^{-\zeta\int^{\f}_{0} X(t)dt},\\
\label{fp} \frac{d\f}{dr} &=&  - \frac{4}{\ell \, \zeta} X(\f) \, e^{-\zeta\int^{\f}_{0} X(t)dt},\\
\label{gp}
\frac{1}{f} \, \frac{df}{dr}
&=& -\frac{4}{\ell}~Y(\f) \, e^{-\zeta\int^{\f}_{0} X(t)dt} \, ,
\eea
and
\bea\label{Xeq}
\frac{dX}{d\f} &=& -\zeta~(1-X^2+Y)\le(1+\frac{1}{2\zeta}\frac{1}{X}\frac{d\log V}{d\f}\ri),\\
\frac{dY}{d\f} &=& -\zeta~(1-X^2+Y)\frac{Y}{X } \, , \label{Yeq} \eea
where $\zeta = \sqrt{\frac{8}{3}}$ and we have introduced the phase variables
\begin{equation}\label{XY1}
  X(\f)\equiv \frac{\zeta}{4}\frac{\Phi'}{A'}\, , \qquad  Y(\Phi)\equiv
  \frac{1}{4}\frac{f'}{f\, A'}\, ,
\end{equation}
which are invariant under radial coordinate transformations \cite{LongThermo}.
Note that the metric functions $A$ and $f$ can be expressed in terms of $X$ and $Y$
by integrating directly (\ref{XY1}),
\bea
A(\f) &=& A(\f_c) + \frac{\zeta}{4}\int_{\f_c}^{\f} \frac{d\tilde{\f}}{X} \, ,\label{Aeq1}\\
f(\f) &=&  \exp\le(\zeta \int_{0}^{\f} \frac{Y}{X} \, d\tilde{\f}\ri) \, , \label{feq1}
\eea
where $\f_c$ denotes a cut-off surface that plays the role of the regularized UV boundary\footnote{
At the boundary we require $f \rightarrow 1$, which fixes its dependence on $\f_c$. Moreover, as shown in
\cite{LongThermo}, the physical observables in this system only depend on $X$ and $Y$ and are independent of $\f_c$.}.

The advantage of expressing Einstein's equations in terms of the phase variables $\{X,Y\}$ should now be apparent --
the system has been reduced to a coupled set of first order equations\footnote{In order to solve
(\ref{Xeq}) and (\ref{Yeq}) one has to specify a single boundary condition for each of the equations.
After demanding a regular horizon of the form
\be\lab{reghor}
f(\f) = const. \times (\f_h-\f) \qquad \text{near}\,\,\, \f \approx \f_h,
\ee
inspection of the equations (\ref{feq1}) and (\ref{Yeq}) shows that the remaining integration constant is completely fixed,
and near $\f\approx \f_h$ one must require
\be\lab{XYh} X(\f) = - \frac{1}{2\zeta} \frac{V'(\f_h)}{V(\f_h)} + \cO(\f_h-\f) \, , \qquad Y(\f) = - \frac{X(\f_h)}{\zeta(\f_h-\f)} +  \cO(1) \, .
\ee
}, (\ref{Xeq}) and (\ref{Yeq}), which greatly simplifies the task of finding analytic solutions.
Moreover, all thermodynamic properties of the system are completely determined
by knowledge of the two degrees of freedom $X$ and $Y$ as a function of $\f$.
In fact, in terms of the horizon value $\f_h$ of the dilaton, the temperature and the entropy of the black brane
are given by
\bea \lab{Teq1} T(\f_h) &=&
\frac{\ell}{12\pi}~e^{A(\f_h)}~V(\f_h)~e^{ \zeta\int_{0}^{\f_h} X(\f)~d\f}\, ,\\
\label{ent31}
S &=& \frac{1}{4 G_N} \, e^{3 A(\f_h)}\, .
\eea
Combining these two expressions we find a simple relation between the phase variable $X$ and the scalar potential,
\be\lab{use1}
e^{-\zeta\int_{0}^{\f_h} X(\f)~d\f} = C~\frac{S^{\frac{1}{3}}}{T} V(\f_h),
\ee
where the proportionality constant $C$ is given by
\be\lab{Cdef}
C = \frac{\ell (4\pi)^{-\frac{4}{3}}}{3 M_p} \, ,
\ee
and we have traded Newton's constant for the Planck mass, $M_p^{3} = (16\pi G_N)^{-1}$.
Finally, the temperature can be directly related to the horizon expansion of the blackness function as follows,
\be\lab{Teq2}
4\pi \, T = - f'(r_h) \, e^{A(r_h)} \, .
\ee
These relations greatly facilitate the study of black brane solutions to the model described by (\ref{fullaction}), as we show in the main text.

\section{Chamblin-Reall from $U(1)$ Reduction}\label{app:U1}

In this appendix we expand upon the particular case of the Chamblin-Reall black brane.
The special case $\gamma = - \alpha$, for which $\etas$ shows no temperature dependence,
can be realized as a solution of a dimensionally reduced theory of pure gravity plus a negative cosmological constant in six dimensions.
As such this solution can be uplifted to a pure AdS black brane in six-dimensions,
thus explaining the temperature independent form for $\etas.$

To see this in more detail, we start from the following Lagrangian density in $d+1$ space-time dimensions,
\begin{equation}
\mathcal{L} =   R + 2\Lambda_{d+1} + \beta R_{\mu\nu\rho\sigma}R^{\mu\nu\rho\sigma} \, ,
\end{equation}
and perform a circle reduction by decomposing the metric as
\begin{equation}
ds_{d+1}^2 = e^{2\sigma\phi}ds_d^2 + e^{2\delta\phi}d\psi^2 \, ,
\end{equation}
where $\phi$ is a scalar\footnote{Note that we have consistently fixed the Kaluza-Klein gauge field to vanish.}.
The dimensionally reduced theory is then given by
\begin{eqnarray}
\mathcal{L}_d &=& R - \sigma^2(d-1)(d-2)(\partial\phi)^2 + 2\Lambda_{d+1}e^{2\sigma\phi} +
\beta e^{-2\sigma\phi}\Big[R^{\alpha\beta\gamma\delta}R_{\alpha\beta\gamma\delta} +  ... \Big] \; ,
\end{eqnarray}
where we fixed $\delta = -(d-2) \, \sigma$ to ensure a canonical Einstein term, and
omitted higher derivative terms involving $R_{\mu\nu}$ and $\phi$, since they do not affect $\etas$.
Notice that, as expected, we have generated a dilatonic coupling to the Riemann-squared term.
However -- unlike in (\ref{fullaction}), where the parameter $\gamma$ is free -- here the dilatonic coupling
is completely fixed in terms of $\alpha$. Thus, we have recovered the model of (\ref{fullaction}) with the CR potential (\ref{VCR}),
but only for the special parameter choice $\gamma = -\alpha$.

We can now compute $\etas$ for the dimensionally reduced theory.
Restricting to $d=5$ and rescaling the dilaton so that its kinetic term agrees with (\ref{fullaction}), the lagrangian becomes
\begin{eqnarray}\label{Lag5}
\mathcal{L}_5 &=& R - 2(\partial\phi)^2 + 2\Lambda_{6}e^{\sqrt{2/3} \phi} +
\beta e^{-\sqrt{2/3} \phi}\Big[R^{\alpha\beta\gamma\delta}R_{\alpha\beta\gamma\delta} + ... \Big].
\end{eqnarray}
Comparing this to our action (\ref{fullaction}), we see that we have generated a scalar coupling
$\sim e^{\gamma\Phi}R_{\mu\nu\rho\sigma} R^{\mu\nu\rho\sigma}$ with $\gamma = -\sqrt{2/3}$.
The black-brane solution to the two-derivative theory is given in \cite{Gouteraux:2011ce},
\begin{eqnarray}
ds_2 &=& e^{2A}(-f(r)dt^2 + d\vec{x}^2) + \frac{dr^2}{f(r)}, \\
e^A &=& r^4, \\
f(r) &=& 1 - \left(\frac{r_h}{r}\right)^{15}, \\
e^{-\sqrt{2/3} \phi} &=& \frac{\Lambda_6 r^2}{90},
\end{eqnarray}
and has the following near horizon expansion
\begin{eqnarray}
f(r)&=& \ft{15}{r_h}(r-r_h) + ... , \nn \\
A(r) &=& \ln{r^4_h} + \ft4{r_h}(r-r_h) + ..., \nn \\
\phi(r) &=& \phi_h - \ft{\sqrt6}{r_h}(r-r_h) + ... \, .
\end{eqnarray}
Plugging this near horizon expansion into our solution for $\etas$ and taking $\Lambda_6 = 10/\ell^2$ we arrive at the following result
\begin{equation}
\etas = \frac{1}{4\pi} \left[1 - 20\beta \right] \, .
\end{equation}
This is precisely the result expected for an $AdS$-Schwarzschild black hole in six-dimensions \cite{Kats:2007mq}.
This is expected since, as we have shown, this specific Einstein-dilaton system in 5-dimensions was simply a
dimensional reduction of a 6-dimensional theory containing only gravity and a cosmological constant. In fact, the dilaton solution presented
may be explicitly uplifted to an $AdS$-Schwarzschild black hole in six-dimensions \cite{Gouteraux:2011ce}. While we have focused on a circle reduction from six-dimensions, similar statements can be made for generic n-torus reductions to five-dimensions as in \cite{Gubser1}, resulting in Chamblin-Reall theories with $\alpha = -\gamma = \sqrt{\frac{8n}{3(n+3)}}.$

\end{document}